\documentclass[12pt]{article}
\usepackage{amsmath, amssymb, amscd, amsthm, amsfonts}
\usepackage{graphicx}
\usepackage{hyperref}
\usepackage{here}
\usepackage{caption}
\usepackage{subcaption}

\usepackage[round]{natbib}

\usepackage{rotating}
\usepackage{tikz}

\usepackage{pdflscape}
\usepackage{bm}
\usepackage{setspace}

\title{Hidden Markov \Polya trees for high-dimensional distributions}
\author{Naoki Awaya and Li Ma \\
\it Department of Statistical Science\\ \it Duke University, Durham, NC 27708, USA}
\date{\today}

\newtheorem{thm}{Theorem}[section]
\newtheorem{lem}{Lemma}[section]
\newtheorem{cor}{Corollary}[section]
\newtheorem{prop}{Proposition}[section]

\usepackage{amsmath}               
  {
      \theoremstyle{plain}
      \newtheorem{assumption}{Assumption}
  }
  
\newcommand{\argmax}{\mathop{\rm arg~max}\limits}
\newcommand{\argmin}{\mathop{\rm arg~min}\limits}

\newcommand{\1}{\mathbf{1}}

\newcommand{\x}{\mathbf{x}}

\newcommand{\Q}{\mathbf{Q}}

\newcommand{\T}{\mathcal{T}}

\newcommand{\Polya}{P\'{o}lya\ }

\newcommand{\bV}{\mathbf{V}}
\newcommand{\bv}{\mathbf{v}}
\newcommand{\sV}{\mathcal{V}}

\newcommand{\bZ}{\mathbf{Z}}
\newcommand{\bz}{\mathbf{z}}

\newcommand{\E}{\mathbb{E}}

\DeclareMathOperator*{\plim}{p-lim}

\newcommand{\sS}{\mathcal{S}}

\newcommand{\Op}{\mathcal{O}_p(1)}

\newcommand{\bxi}{{\bm \xi}}
\newcommand{\bgamma}{{\bm \gamma}}

\newcommand{\iid}{\stackrel{\mathrm{iid}}{\sim}}
\newcommand{\ind}{\stackrel{\mathrm{ind}}{\sim}}

\newcommand{\bx}{{\bm x}}
\newcommand{\bthe}{{\bm \theta}}

\usepackage{algorithm}
\usepackage{algpseudocode}

\usepackage{bibspacing}
\begin{document}
\maketitle

\begin{abstract}
The P\'olya tree (PT) process is a general-purpose Bayesian nonparametric model that has found wide application in a range of inference problems. It has a simple analytic form and the posterior computation boils down to beta-binomial conjugate updates along a partition tree over the sample space. Recent development in PT models shows that performance of these models can be substantially improved by (i) allowing the partition tree to adapt to the structure of the underlying distributions and (ii) incorporating latent state variables that characterize local features of the underlying distributions. However, important limitations of the PT remain, including (i) the sensitivity in the posterior inference with respect to the choice of the partition tree, and (ii) the lack of scalability with respect to dimensionality of the sample space. We consider a modeling strategy for PT models that incorporates a flexible prior on the partition tree along with latent states with Markov dependency. We introduce a hybrid algorithm combining sequential Monte Carlo (SMC) and recursive message passing for posterior sampling that can scale up to 100 dimensions. \textcolor{black}{While our description of the algorithm assumes a single computer environment, it has the potential to be implemented on distributed systems to further enhance the scalability.} Moreover, we investigate the large sample properties of the tree structures and latent states under the posterior model. We carry out extensive numerical experiments in density estimation and two-group comparison, which show that flexible partitioning can substantially improve the performance of PT models in both inference tasks. We demonstrate an application to a mass cytometry data set with 19 dimensions and over 200,000 observations. 
\end{abstract}

\newpage 

\doublespacing

\section{Introduction}
\label{sec: Introduction}
The \Polya tree (PT) \citep{freedman1963asymptotic, ferguson1974prior,lavine1992some} is a stochastic process that generates random probability measures and is introduced as a prior for Bayesian nonparametric inference. While the PT generalizes the Dirichlet process (DP) \citep{ferguson1973bayesian} as it yields the DP under certain hyperparameters \citep{ferguson1974prior}, the statistical properties and practical applications of the PT are very different. While the DP is most frequently used as a mixing distribution that induces clustering structures, the PT is often adopted for directly modeling probability densities. 

The PT is defined generatively on a recursive partition---or a partition tree---over the sample space through coarse-to-fine sequential probability assignment at each tree split. 
In a classical (univariate) PT, the tree is dyadic and the conditional probability assigned to the two children nodes at each tree split arises from independent beta priors, which leads to analytic simplicity and ease in computing the posterior. Obtaining the posterior is straightforward from beta-binomial conjugacy and incurs a computational budget that scales only linearly with the sample size, \textcolor{black}{making the PT one of the few nonparametric models applicable to data with massive sample size. Moreover, the posterior computation is embassingly parallelizable across the tree nodes}.

The PT has been applied in various contexts beyond the original application of density estimation. A far-from-exhaustive list includes survival analysis \citep{muliere1997bayesian, neath2003polya}, imputing missing values \citep{paddock2002bayesian}, goodness-of-fit tests \citep{berger2001bayesian}, two-group comparison \citep{ma2011coupling, holmes2015two, chen2014bayesian, soriano2017probabilistic}, density regression \citep{jara2011class}, ANOVA  \citep{ma2018analysis}, testing independence \citep{filippi2017bayesian}, and hierarchical modeling \citep{christensen2019bayesian}. The PT has also been utilized in semi-parametric analyses such as in (generalized) linear models \citep{walker1999bayesian,hanson2002modeling,walker1997hierarchical}.

Early developments of the PT are based on an {\em a priori} fixed partition tree on the sample space. The resulting inference can be sensitive to the choice of the partition points defining the tree. In particular, the resulting process, both {\em a priori} and {\em a posteriori}, can be jumpy at these points. In hypothesis testing and model choice, this sensitivity is also reflected in the sometimes substantial change in the marginal likelihood/Bayes factor when the partition points are slightly varied. To remedy the issue, \citet{paddock2003randomized} modified the PT model so that  observations are generated from the PT model with slightly different partition points. \citet{hanson2002modeling} and \citet{hanson2006inference} proposed a mixture of PTs by defining partition points along quantiles of a parametric model endowed with a hyperprior to allow model averaging on the partition points. This strategy does not allow individual partition points to adapt to local features of the distribution but only the whole set of points to the global structure of the distribution, and is most effective when the underlying density is close to the specified parametric model. \citet{nieto2012rubbery} took a different approach by modeling the probability assignments within each level of the tree in a correlated manner to smooth out the random measure over the boundaries of partitioning. While these approaches alleviate the sensitivity to partition points in low-dimensional settings, they are not easily applicable (though in principle possible) to problems with even just a handful of dimensions. Moreover, Bayesian inference with these models generally require MCMC, whose effectiveness can (in fact often does) still suffer from the sensitivity with respect to the partition points.

\textcolor{black}{Another related issue regarding the partitioning scheme of the PT is that in multivariate problems, traditionally the partition tree is constructed by dividing all dimensions of the sample space at each split. For example, for a $d$-dimensional sample space, each time a tree node is divided, it is split into $2^{d}$ children nodes, and probability assignment over these $2^{d}$ child nodes is modeled by independent Dirichlet priors.} \cite{wong2010optional} noted that  such a ``symmetric'' partition scheme is undesirable as the dimensionality increases, in which case due to the exponential growth of the partition blocks, the vast majority of the blocks are barely, if at all, populated by data. As such they propose to incorporate adaptivity into the partitioning strategy with respect to the structure of the underlying distribution through adopting a Bayesian CART-like prior \textcolor{black}{\citep{chipman1998bayesian} on the space of dyadic partition trees.}

However, in order to maintain the analytic simplicity of the posterior and achieving MCMC-free exact Bayesian inference with a linear computational budget, 
the Bayesian CART-like prior has to be restricted to only divide at the middle point (or otherwise a pre-determined fixed point) on one of the dimensions at each tree split. Not only does this hamper the model's ability to adapt to distributional structures, but it makes the model suffer from the same sensitivity with respect to the partition points. Moreover, even with this restriction, the inference algorithm (based on recursive message passing) is only computationally practical for up to about 10 dimensions on continuous sample spaces.

In a different vein, recent developments have demonstrated that aside from enhancing the partitioning strategy, the PT can also be substantially improved by adopting more flexible priors (as opposed to independent betas) on the probability assignment at each tree split \citep{jara2011class,nieto2012rubbery,ma2017adaptive}. One strategy for enriching the PT in this regard is by introducing latent state variables at each tree split and adopt priors on the probability assignment {\em given} these states. 
When the latent states are discrete and modeled with Markov dependency, analytical simplicity is preserved and exact Bayesian inference can proceed through recursive message passing that maintains the linear computational budget \citep{ma2017adaptive}.

Given these developments, we are motivated by the following questions: Is it possible to incorporate into the PT a very flexible partition tree prior, such as the general Bayesian CART (i.e., without the restriction to partition at middle points), that will (i) enhance its adaptivity to distributional structures in multivariate settings; (ii) resolve its sensitivity to the choice of partition points; and (iii) allow a tractable form of the joint posterior and a posterior inference algorithm that is scalable to moderately high-dimensional problems (e.g., up to 100 dimensions)? Moreover, should such a strategy exist, can the resulting model and inference algorithm be made compatible with incorporating (possibly Markov dependent) latent states on the tree nodes?

The goals of making the partition tree prior more flexible while enhancing the computational scalability appear at odds with each other. Large tree spaces are well known to be very hard to compute over. In moderate to high dimensional settings exact inference involving flexible tree structures is beyond reach and even the most advanced MCMC approaches tailor-made for trees encounter substantial difficulty due to the pervasive multi-modality of distributions in such spaces. 
Recent advances in sequential Monte Carlo (SMC) for regression tree models \citep{lakshminarayanan2013top,lu2013multivariate}, however, suggest that efficient inference is possible in moderately high-dimensional settings (up to about 100 dimensions). 
Moreover, once the partition tree is sampled, the conditional posterior for the rest of the model can be computed analytically through recursive message passing. 
We will therefore exploit a hybrid strategy that uses a new SMC sampler to efficiently sample from the marginal posterior of the partition tree structure, \textcolor{black}{along with} recursive message passing to compute the exact conditional posterior of the latent state variables given the tree.

Beyond the methodological development, we will also investigate the theoretical properties of the posterior on the partition tree and the latent states. Previous theoretical literature on the PT and related models have mostly focused on establishing the posterior consistency and the contraction rate of the random measure induced under these models \citep{castillo2017polya,castillo2021optional}.  
In multivariate settings, however, the partition tree itself is highly informative about the underlying distribution. Moreover, in applications involving model choice and hypothesis testing, it is often the latent states, not the random measures, that are of direct interest. As such, we focus on studying the asymptotic behavior of the marginal posterior on the partition tree and  latent states, establishing consistency results on their convergence toward the trees and states that most closely characterize the underlying truth.

The rest of the paper is organized as follows. In Section~\ref{sec: method} we describe a flexible prior on the partition tree structure that relaxes the restriction of ``dividing in the middle'' on partition points and present a general form of PT models that adopt this prior along with latent states associated with the tree nodes with a Markov dependency structure. 
In Section~\ref{sec: posterior inference}, we present our hybrid computational strategy that can work effectively up to 100 dimensions consisting of an SMC algorithm for sampling on the marginal posterior of the partition tree and a recursive message passing algorithm for obtaining the exact conditional posterior of the latent states and the random measure given the sampled trees.
In Section~\ref{sec: theory} we investigate the asymptotic properties of the tree structures and latent states identified under the posterior model.
In Section~\ref{sec: experiments}, we carry out extensive numerical experiments to examine the performance of our method in the context of two important applications of PTs---density estimation and the two-group comparison, followed by an application to a data set from mass cytometry in Section~\ref{sec:mass}.
In Section~\ref{sec: conclusion} we conclude with a brief discussion.
\textcolor{black}{
All proofs are provided in {\bf Supplementary Materials~C}.
}

\vspace{-1em}
\section{Method}
\vspace{-0.5em}
\label{sec: method}

We first review the PT model \citep{ferguson1973bayesian,lavine1992some} \textcolor{black}{on a dyadic recursive partition} in Section~\ref{sec: plain vanilla PT}. \textcolor{black}{The model, while defined on a general multivariate sample space, differs from a traditional multivariate PT which adopts a multi-way symmetric recursive partitioning.}
Then we introduce a new class of PT models that incorporates both the flexible partition prior and latent states with Markov dependency.

\vspace{-1em}
\subsection{\Polya trees \textcolor{black}{defined on recursive dyadic partitions}}
\label{sec: plain vanilla PT}
\vspace{-0.5em}

Without loss of generality, we consider a continuous sample space represented as a $d$-dimensional rectangle $\Omega = (0,1]^d$.
\textcolor{black}{For unbounded sample spaces such as $\mathbb{R}^d$, one can transform each margin to [0,1] by applying, say, a cumulative distribution function transform or by standardizing the data based on its observed range of values.}
 We use $\mu$ to denote the Lebesgue measure on $\Omega$.
A (dyadic) recursive partitioning $T$ on $\Omega$ is a sequence of partitions of $\Omega$ such that the partition blocks at each level of the partitioning are obtained by dividing each block in the previous level into two children blocks. Formally, we can write $T = \bigcup^\infty_{k=0} \mathcal{A}^k$, where $\mathcal{A}^k$ is a partition of $\Omega$ in the $k$th level.
More specifically, $\mathcal{A}^0 = \{
\Omega\}$, and $A \in \mathcal{A}^k$ ($k=0, 1,2,\dots$) is divided into $A_l$ and $A_r$, which satisfy $A_l, A_r \in \mathcal{A}^{k+1}$, $A_l \cup A_r = A$, and $A_l \cap A_r = \emptyset$. 
(Throughout the paper, a subscript ``$l$'' or ``$r$'' on a node indicates the left or right child node.)
For example, when $d = 1$ and if the tree is recursively divided at the middle point of each node, then nodes in level~$k$ are of the form $(l/2^k, (l+1)/2^k]$ for some $l\in \{0,\dots,2^k-1\}$.
Another common strategy is to define the tree based on the quantiles of a probability measure $F$ so that $A \in \mathcal{A}^k$ is of the form $A=(F^{-1}(\frac{l}{2^k}), F^{-1}(\frac{l+1}{2^k})]$ for $l\in 0,\dots,2^k-1$.

Given a partition tree $T$, we can define a random measure $Q$ by putting a prior on the conditional probability $\theta(A)= Q(A_l| A) = 1 - Q(A_r| A)$ at each $A \in T$.
Under the PT prior, the parameters $\theta(A)$ follow independent beta distributions Beta$(\alpha_l(A), \alpha_r(A))$, where $\alpha_l(A)$ and $\alpha_r(A)$ are hyperparameters.
\textcolor{black}{The corresponding posterior, given an i.i.d.\ sample $x_1,\ldots,x_n$ from $Q$, is again a PT} with a simple conjugate update on the conditional proabilities:
\vspace{-1.5em}
 \[
     \theta(A) \mid x_1,\dots, x_n
     \sim {\rm Beta}(\alpha_l(A) + n(A_l), \alpha_r(A) + n(A_r)),
 \]
 \vspace{-3.5em}
 
\noindent 
where $n(A)$ represents the number of observations in a set $A\subset \Omega$.
Though the tree needs to be infinitely deep to ensure full support of the PT, for practical purposes, one typically sets a sufficiently large maximum depth (or resolution) of $T$ and compute the posteriors of $\theta(A)$'s defined on this finite tree structure \citep{hanson2002modeling}.
We shall refer to a node in the deepest level as a ``leaf'' or ``terminal node''. On a leaf, the conditional distribution can be set to a baseline \textcolor{black}{$F(\cdot|A)$}, such as the uniform distribution $\mu(\cdot| A)$. 
In Section~\ref{sec: posterior inference} when we present inference algorithms, we shall adopt this practical strategy and assume $T$ is finite and use $\mathcal{N}(T)$ and $\mathcal{L}(T)$ to denote the collection of the non-terminal nodes and the leaf nodes, respectively. 
\vspace{-1em}
 \subsection{Incorporating flexible partition points}
 \label{sec: randomizing tree}
\vspace{-0.5em}

 We incorporate a Bayesian-CART like prior on $T$ by randomizing both the dimension in which to divide a node and the location to divide. Our prior relaxes the ``always-divide-in-the-middle'' restriction imposed in \cite{wong2010optional}. \textcolor{black}{This prior on the partition tree $T$ differs from that in the mixture of PTs of \cite{hanson2006inference}, which does not randomize over the dimension to divide, but generates the boundaries of the tree nodes jointly using quantiles of a parametric family.}
 
 \textcolor{black}{Our prior can be described iteratively as a generative process that recursively divides the sample space. Specifically,}  suppose we have a node $A$ in the rectanglar form,
     $A = (a_1,b_1] \times \cdots \times (a_d, b_d]$.
 We divide $A$ into two rectangular children by cutting along a randomly chosen dimension at a random location. The dimension to divide $D(A)\in\{1,2,\ldots,d\}$, and the (relative) location to divide $L(A)\in(0,1)$ are given independent priors of the following forms:
\vspace{-1.5em}
 \begin{align}
    D(A) &\sim \mathrm{Mult}(\lambda_1(A), \dots, \lambda_d(A)) \quad \text{and} \quad  L(A) \sim \sum^{N_L-1}_{l=1} \beta_l(A) \delta_{l/N_L} (\cdot), 
    \label{prior of d and l}
\end{align}
 \vspace{-3.5em}
 
\noindent 
where $\delta_{x}(\cdot)$ represents the unit point mass at $x$, and $N_L - 1$ is the total number of grid points along $(0,1)$. Both \textcolor{black}{$\{\lambda_i(A)\}_{i=1,\dots,d}$ } and $\{\beta_l(A)\}_{l=1,\dots,N_L-1}$ sum to 1. 
In the above, we have adopted a uniform grid over $(0,1)$ \textcolor{black}{for notational simplicity}, but it does not have to be as such.  With $D(A) = j$ and $L(A) = l/N_L$, the two children nodes $A_l$ and $A_r$ are 
\vspace{-1.5em}
 \begin{align*}
     A_l &= (a_1,b_1] \times \cdots \times (a_j, a_j + {l}/{N_L} \cdot (b_j - a_j)]
     \times \cdots \times (a_d, b_d], \\
     A_r &= (a_1,b_1] \times \cdots \times (a_j + {l}/{N_L} \cdot (b_j - a_j), b_j]
     \times \cdots \times (a_d, b_d].
 \end{align*}
 \vspace{-3.5em}

In principle one could adopt a continuous prior on the partition location $L(A)$. A discretized prior is helpful, however, because it \textcolor{black}{will substantially simplify} posterior computation. In practice, as long as the grid is dense enough, the discrete prior will be practically just as flexible.  
Indeed we have verified in extensive numerical experiments that when $N_L$ is large enough (more than 30 to 50) over a uniform grid, posterior inference no longer improves in any noticeable way. 

For the prior on $D(A)$, we set $\lambda_j(A) = 1/d$ for all nodes $A$ as a default choice. 
When $L(A)$ is given a weak prior widely spread over $(0,1)$, the resulting inference can be sensitive to the ``tail'' behavior of the distribution in the node, 
\textcolor{black}{
resulting in high posteriors of $L(A)$ near the extreme values 0 and 1.
A detailed discussion on this phenomenon will be provided in Section~\ref{subsubsec: 2D density estimation}.} 
This issue can be effectively addressed by making the prior of $L(A)$ depend on the sample size $n(A)$ so that it encourages more balanced divisions at large sample sizes.  More specifically, we adopt the following prior with an exponentially decaying tail
\vspace{-1.5em}
\begin{align}
    P(L(A) = l/N_L)
    =
    \beta_l
    &\propto
    \exp
    \left[
        - \eta n(A) f(|l/N_L - 0.5|)
    \right],\ 
    l = 1,\dots,N_L-1,
    \label{prior of L}
 \end{align}
 \vspace{-3.5em}
 
 \noindent
where $\eta \geq 0$ is a hyperparameter and \textcolor{black}{$f:[0,\infty)\rightarrow [0,\infty)$} is an increasing function \textcolor{black}{with $f(0) = 0$}. In the following, we shall use a function $f(x) = x$, and so our prior on $L(A)$ is a (discretized) Laplace distribution. \textcolor{black}{We provide theoretical justification for adopting this prior with exponential tails in Section~\ref{sec: theory}.}

Another generalization of the prior on $L(A)$ is to incorporate a spike-and-slab set-up with a spike at the middle point $1/2$. In particular, one can adopt a dependent spike prior among the nodes such that once a node $A$ is divided exactly at the middle point, so are its descendants. This generalization will substantially reduce the amount of computation in regions of the sample space where the data are either sparse or lack interesting structure, e.g., close to the uniform distribution. We implement the spike-and-slab in our software but defer the details of this generalization to {\bf Supplementary Materials~A} to avoid distracting the reader from the main ideas.

\begin{figure}[htb]
    \begin{subfigure}[b]{0.33\textwidth}
         \centering
        \includegraphics[width=0.4\linewidth]{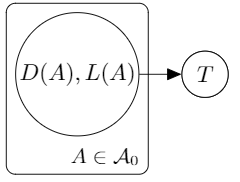}
        \caption{Partition tree generation\\
        ($\mathcal{A}_0$: All potential nodes)}
    \end{subfigure}%
    \begin{subfigure}[b]{0.33\textwidth}
         \centering
        \includegraphics[width=0.6\linewidth]{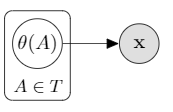}
        \caption{PT without latent states}
    \end{subfigure}%
    \begin{subfigure}[b]{0.33\textwidth}
         \centering
        \includegraphics[width=0.7\linewidth]{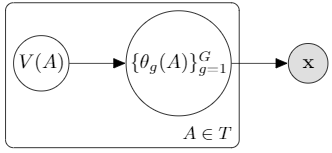}
        \caption{PT with latent states}
    \end{subfigure}%
\caption{Graphical representation of PT models given the tree $T$. The hyperparameters are hidden for simplicity.}
\label{fig: HMPT graphical representation}
\end{figure}

Given the tree prior, our PT model now consists of the two components---tree generation and conditional probability assignment. Figures \ref{fig: HMPT graphical representation}(a) and \ref{fig: HMPT graphical representation}(b) present a graphical model representation for each.

\vspace{-2em}
\subsection{Hidden Markov \Polya tree models}
\label{sec: HMM framework}
\vspace{-0.5em}
\subsubsection{General framework}
\label{sec: HMM General framework}
\vspace{-0.5em}
Next we extend the above model to accommodate two recent developments in the PT literature: (i) incorporating latent state variables along the tree structure and (ii) joint modeling of \textcolor{black}{multiple groups of observations}. 
\textcolor{black}{Incorporating latent variables allow more flexibly characterizing distributional features through adding prior dependency.} As in recent literature, we consider incorporating discrete state variables that follow a Markov process along the tree structure. Because the description in this section always pertains to the model {\em given} the randomly generated partition tree~$T$, for brevity we shall not keep stating ``given $T$''.

\textcolor{black}{
 We generalize our notation to allow observing one or more groups of i.i.d.\ observations. Let $G$ be the number of groups of i.i.d.\ observations. For the $g$th group ($g=1,2,\ldots,G$), let $Q_g$ be the sampling measure for that group. Let $\Q$ denote the collection of all $G$ sampling measures. That is, $\Q = \{Q_g\}^G_{g=1}$. 
Let $\bx_g = (x_{g,1},\dots,x_{g,n_g})$ denote the observations in the $g$th group, which are i.i.d.\ given $Q_g$, where $n_g$, the sample size for the group, is allowed to differ across the groups. We use $\x = \{\bx_g\}^G_{g=1}$ to denote the collection of all observations from all groups.
}

Next we specify a prior on $\Q$ in terms of a joint prior on the conditional probability on each $A\in T$,
    $\theta_g(A) = Q_g(A_l \mid A) = 1-Q_g(A_r\mid A)$.
We use latent variable modeling to incorporate \textcolor{black}{prior} dependency among the tree nodes. Specifically, let $\{V(A): A \in T\}$  denote a collection of latent state variables, one for each $A$, and without loss of generality, assume that $V(A)$ takes discrete values from $\{1,\dots,I\}$. (In practice, the number of states $I$ can differ among $A$.) Joint priors of $\theta_g(A)$ for all $g$ and $A$ are then defined conditionally on these latent states. 

Existing literature has exploited these latent states to characterize both the within-\textcolor{black}{group} structure of each distribution $Q_g$ and the between-\textcolor{black}{group} relationship among the $Q_g$. An example of within-sample structures is the smoothness of each underlying distribution, which is explored in the context of density estimation \citep{ma2017adaptive}. An example of between-\textcolor{black}{group} structures is the difference between two (or more) distributions \citep{soriano2017probabilistic}. 

Dependent modeling of the latent states over the partition tree is desirable as {\em a priori} one would expect interesting structures (both within-\textcolor{black}{group} and between-\textcolor{black}{group}) to exhibit themselves in a correlated manner over the sample space. \textcolor{black}{For example}, functions tend to have similar smoothness over adjacent locations, and two-\textcolor{black}{group} difference tend to be clustered in space. A computationally efficient strategy for modeling such dependency over the tree is by a hidden Markov process along the tree \citep{crouse1997contextual}, which starts from the root node, \textcolor{black}{$A=\Omega$}, and sequentially generates the latent states in a coarse-to-fine fashion according to (possibly node-specific) transition matrices ${\bm \bxi}(A)$ whose $(i,i')$th element is
 \vspace{-1.5em}
 \begin{align*}
 \bxi_{i,i'}(A)
 &= 
 P(V(A) = i' \mid V(A^p) = i)
  \quad \text{for } i,i'\in\{1,\dots,I\},
 \end{align*}
  \vspace{-3.5em}
 
 \noindent
 where $A^p$ denotes $A$'s parent.
 (We shall use superscript ``$p$'' to indicate the parent of a node in $T$.) \textcolor{black}{For $A = \Omega$, since $\Omega$ has no parent, we can simply let $\bxi_{i,i'}(\Omega)$ be constant over $i$, representing the initial state probabilities on $\Omega$.}
 
 Given the $V(A)$'s, $\{\theta_g(A)\}_{g=1,\dots,G}$ can then be modeled as conditionally independent \textcolor{black}{{\em a priori}}. The specific choices of these conditional priors are problem-dependent. We give two examples below. Figure~\ref{fig: HMPT graphical representation}(c) presents a graphical model representation for the latent state modeling \textcolor{black}{on $G$ probability distributions by PTs} given $T$, which along with our generalized prior on the partition tree $T$ presented in Figure \ref{fig: HMPT graphical representation}(a) forms the most general version of the model we consider in this work.
 \vspace{0.5em}
 
  \noindent {\bf Example 1: Density estimation with adaptive smoothness}
 \label{subsec: examples}
 
 \noindent An example of within-\textcolor{black}{group} structures that the latent state $V(A)$ can characterize is the smoothness of the density functions for the random measures. 
 For example, \cite{ma2017adaptive} proposed the adaptive \Polya tree (APT) model which incorporates latent states to allow different levels of local smoothness in the underlying distribution. This is achieved by modeling the $\theta(A)$'s as Beta$(m(A)\nu(A), (1-m(A))\nu(A))$, where $m(A)$ is the prior mean and $\nu(A)$ the precision parameter which characterizes the smoothness of the random measure with larger $\nu(A)$ corresponding to more smoothness, and then modeling the precision parameters conditional on the latent state $V(A)$ with a hyperprior
$ \nu(A)\mid V(A)=i \sim F_i $
with the (conditional) hyperprior $F_i$ given $V(A)=i$. \textcolor{black}{The hyperpriors $F_1,F_2,\ldots,F_I$ are chosen to be stochastically increasing $F_1 \prec F_2 \prec \cdots \prec F_I$. That is,  $F_i([x,\infty)) \leq F_{i+1}([x,\infty))$ for all $x >0$. The rate at which the latent states transition into a higher latent state along each subbranch of the partition tree determines the local smoothness of the underlying density.}
\vspace{1em}

\noindent
{\bf Example 2: \textcolor{black}{Two-group comparison}}\\
In two-group comparison, we are interested in testing and identifying differences between two measures $\Q = \{Q_g\}_{g=1,2}$ based on an i.i.d.\ sample from each. The ``global'' testing problem can be formulated as testing the following null and alternative hypotheses:
$H_0: Q_1 = Q_2$ vs $H_1:Q_1 \neq Q_2$. 
Noting that two-group differences may exist in parts of the sample space and not others, the coupling OPT \citep{ma2011coupling} and the multi-resolution scanning (MRS) model \citep{soriano2017probabilistic} are PT-based models that allow the measures to differ on some nodes $A\in T$ and not others. This more ``local'' perspective on two-group comparison enables these models to not only test for $H_0$ vs $H_1$, but to identify regions on which the two measures differ. To achieve this, these models incorporate state variables that characterize whether the conditional probabilities on each $A$ are equal: 
\vspace{-1.5em}
\begin{align}
    V(A) = 1  \Leftrightarrow  Q_1(A_l \mid A) \neq Q_2(A_l \mid A), \
    V(A) = 2  \Leftrightarrow  Q_1(A_l \mid A) = Q_2(A_l \mid A).
    \label{eq:2sample_state}
\end{align}
 \vspace{-3.5em}
 
 \noindent
When $V(A) = 1$, the two corresponding conditional probabilities are given independent beta priors, whereas if $V(A) = 2$, they are tied and given a single beta prior. Markov dependency among the states on different nodes are incorporated to induce the desired spatial correlation of \textcolor{black}{cross-group} differences.
Additional latent states can be further incorporated to reflect more complex relationships between the distributions. In fact, the MRS also incorporates an additional state $V(A)=3$, which introduces the same coupled prior as $V(A)=2$, but works as an absorbing state that once $V(A)=3$, all descendants of $A$ will remain in that state, corresponding to the case that the conditional distributions $Q_1(\cdot|A)$, and $Q_2(\cdot|A)$ are completely equal.

\vspace{-1.5em}
\section{Bayesian inference}
\vspace{-0.5em}
\label{sec: posterior inference}
In sum, the models we consider \textcolor{black}{all share a common structure} consisting of the following components: (i) the partition tree $T$ defined by the dimension and location variables $D$'s and $L$'s, which follow the priors given in Eq.~\eqref{prior of d and l};
(ii) the latent state variables $V(A)$ given $T$ which follow a Markov prior; (iii) the conditional probabilities along the given tree $T$, $\{\theta_g(A)\}^G_{g=1}$, whose joint prior are specified independently across the nodes on $T$ given the latent states; and finally (iv) given the random measures $Q_g$ defined by $T$ and $\theta_g(A)$'s, we observe an i.i.d.\ sample $\bx_g$ from each $Q_g$, independently across $g$. 
Formally, we have the following full hierarchical model:
\vspace{-1.5em}
\begin{align*}
    T\mid {\bm \lambda},\eta &\sim p(T\mid {\bm \lambda},\eta)\\
    \{V(A): A\in T\}\mid \bxi,T &\sim {\rm Markov}({\bxi})\\
       (\theta_1(A),\dots,\theta_G(A)) \mid V(A), T
    &\ind p(\theta_1(A),\dots,\theta_G(A) \mid V(A)) \text{ for $A\in T$}\\
    \bx_{g}=(x_{g,1},x_{g,2},\ldots,x_{g,n_g})\mid Q_{g} &\iid Q_{g} \text{ for $g=1,2,\ldots,G$.}
\end{align*}
 \vspace{-3.5em}
 
 \noindent
The key to Bayesian inference is the ability to either compute or sample from the joint posterior $(T,\bV,\bthe)$ given all data $\bx=(\bx_1,\ldots,\bx_G)$, where $\bV$ and $\bthe$ represent the totality of all latent states and conditional probabilities given $T$ respectively. While in some problems such as density estimation one may mainly be interested in just the marginal posterior of the $Q_g$'s, in others such as two-group comparison where one wants to characterize the between-group relationships among the distributions, the latent states (along with $T$), \textcolor{black}{which characterizes such relationhips,} are often of prime interest. In multivariate and even high-dimensional problems, the tree structure $T$ is also of great interest as it sheds light on the underlying structures in the distributions.

To this end, we shall take advantage of recent developments in  sequential Monte Carlo (SMC) sampling for tree-based models \citep{lakshminarayanan2013top,lu2013multivariate} and advances in message passing algorithms for PT models with Markov dependency \citep{ma2017adaptive}. We introduce a hybrid algorithm that combines these two computational strategies to effectively sample from the joint posterior in high-dimensional spaces. Overall, the hybrid algorithm consists of two stages:
\vspace{-1em}

\begin{description}
\item[1. Sampling from the marginal posterior of the partition tree]~\\
 We design an SMC sampler---that is, a particle filter---to sample a collection of tree structures $T^1,\dots,T^M$ by growing each tree from coarse to fine scales. It uses one-step look-ahead message passing to construct proposal distributions for $D(A)$ and $L(A)$, one node at a time. \vspace{-0.5em}
\item[2. Computing the conditional posterior given the sampled trees]~\\
    Given each tree sampled by the SMC, we analytically compute the exact conditional posteriors of $V(A)$'s and $\theta(A)$'s \textcolor{black}{using recursive message passing. (The detailed algorithm will be given in Section \ref{subsec: Posterior computation given sampled tree structures}.)} 
    \vspace{-1.0em}
\end{description}

 \vspace{-1em}
\subsection{SMC to sample from tree posterior}
\label{subsec: smc algorithm}

 In the SMC stage to sample the trees, each particle stores a realized form of a finite tree structure, and one node of each tree is divided at each step of the SMC sampling.
 Suppose $T_t$ is the finite tree obtained after dividing the sample space $t$ times in a particle, and for this tree we define the target distribution
\vspace{-1.5em}
 \begin{align*}
     \pi_t(T_t) = P(T_t \mid \x) \propto P(T_t)P(\x \mid T_t).
 \end{align*}
  \vspace{-3.5em}
 
\begin{figure}[htb]
\centering
\begin{tabular}{c}
    \includegraphics[height=3.5cm]{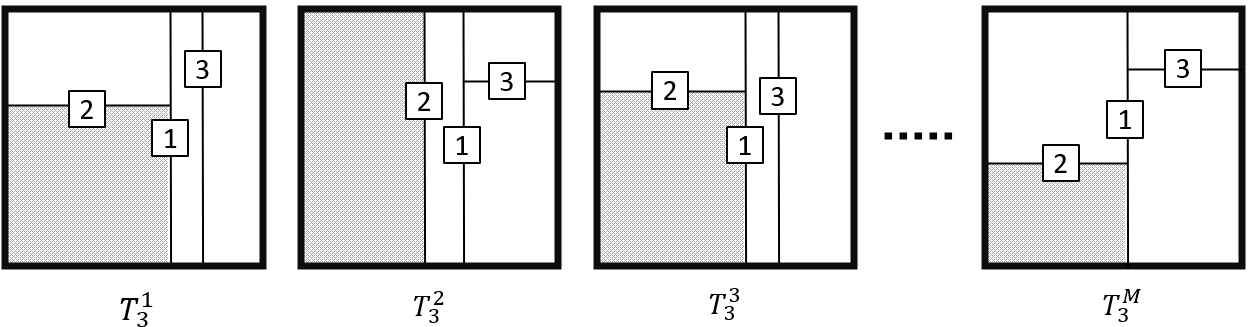}
\end{tabular}
\vspace{-0.5em}
\caption{ 
\textcolor{black}{
An example of realized finite trees in the particle system obtained after the step $t=3$. The numbers in the squares indicate in which step the boundaries are drawn. 
Among the current leaf nodes, the nodes colored gray are the oldest nodes generated in the earliest step, so they are split in the next step. }
}
\label{fig: SMC}
\end{figure}

 \noindent
Here $P(T_t)$ is the joint prior of the variables $D(A)$'s and $L(A)$'s for the non-leaf nodes of $T_t$, and $P(\x \mid T_t)$ is the marginal likelihood given the tree $T_t$ under the hierarchical model, in which $\bV$ and ${\bm \theta}$ are integrated out.
To sample from this target distribution, we sequentially construct a set of $M$ particles $\{T^m_t, W^m_t\}^M_{m=1}$, where $T^m_t$ is a realized tree and $W^m_t$ is the associated importance weight for the $m$th particle. 
 Examples of generated trees are given in Figure \ref{fig: SMC}, where
 the sample space has been divided three times, and in the next step, new partition boundaries will be added in the gray nodes.

 \textcolor{black}{
 Following \cite{lakshminarayanan2013top}, we adopt in each step of the SMC a breadth-first tree-growth strategy by dividing the ``oldest active'' leaf node---that is, the one generated in the earliest step and is yet to be terminated in division. Further division of a node is terminated once the number of observations in that node is below a pre-set threshold (e.g., 5 in our software implementation) to avoid overfitting. 
 }
 Otherwise a node is bisected along a boundary whose dimension and location are randomly drawn from a proposal distribution. For each particle, a finite tree $T_t$ is formed by a sequence of decisions $\{J_s\}^t_{s=1}$, where $J_s = (D_s, L_s)$ correspond to all of the variables $D(A)$ and $L(A)$ at the $s$th step of the SMC.

\textcolor{black}{
As we will see in Proposition \ref{prop: components to make SMC}, the target distribution $\pi_t (T_t)$ has a decomposition $\pi_t(T_t) 
      =
      C_t
     \pi_{t-1}(T_{t-1})
     \pi_t(J_t \mid T_{t-1}) w_t(T_{t-1})$, 
where $C_t$ is a constant independent of $T_t$, $\pi_t(J_t \mid T_{t-1})$ a conditional distribution on $J_t$ given $T_{t-1}$, and $w_t(T_{t-1})$ a function of $T_{t-1}$.
We will choose $\pi_t(J_t \mid T_{t-1})$ as the proposal for $J_t$ under which the corresponding importance weight will simply be $w_t(T_{t-1})$, independent of $J_t$.
}

More specifically, suppose at the current step $t$, we are to divide $A_t \in T_{t-1}$, into $A_{t,l}$ and $A_{t,r}$ with decision $J_t$.
 Let $M_i (A_t \mid J_t)$ be the marginal likelihood on the node $A_t$ under the decision $J_t$ evaluated based on the observations in $A_t$. That is,
 \vspace{-1.0em}
\begin{align}
   \hspace{-1em} M_i (A_t \mid J_t)
    &=\!\!
    \int
    \left[
    \prod^G_{g=1} \theta_g(A_t)^{n_g(A_{t,l})} (1 - \theta_g(A_t))^{n_g(A_{t,r})}
    \right]
    d P(\theta_1(A_t),\dots, \theta_G(A_t) | V(A_t) = i),
    \label{def of MA}
\end{align}
  \vspace{-3.0em}
  
  \noindent
where $n_g(A)$ is the number of observations of the $g$th group included in $A$. \textcolor{black}{To avoid cumbersome notation, we suppress in our notation the dependency of $M_i (A_t \mid J_t)$ on the observations $\x$.}
For example, if the $\{\theta_g(A_t)\}_{g=1}^{G}$ follow independent beta priors written as 
Beta$(\alpha^i_l(A_t), \alpha^i_r(A_t))$ given $V(A_t)=i$, then the marginal likelihood has the following expression $     M_i(A_t \mid J_t)= \prod_{g=1}^{G}
     \frac{B(\alpha^i_l(A) + n_g(A_{t,l}), \alpha^i_r(A) + n_g(A_{t,r}))}{B(\alpha^i_l(A_t), \alpha^i_r(A_t))},$
 where $B(\cdot,\cdot)$ is the beta function. 
 \textcolor{black}{
    Based on the values of $M_i(A_t \mid J_t)$, we can analytically compute the proposal and the importance weight using a general recursive algorithm, as described in the following proposition. 
}
    
\begin{prop}
\label{prop: components to make SMC}
    For every possible decision $J_t$ and states $i=1,\dots,I$, let ${\bf \varphi}_i(A_t)$ be a function defined recursively:
     \vspace{-1.0em}
     \begin{align}
     {\bm \varphi}_i(A_t)
     =
     \begin{cases}
     \frac{\bxi_{1, i}(\Omega)
     M_i(\Omega \mid J_t)}{\sum^I_{j=1} \bxi_{1, j}(\Omega)
     M_j(\Omega \mid J_t)}
     & \text{if $A_t = \Omega$}\\
      \frac{\sum^I_{j=1} {\bf \varphi}_j(A^p_t) \bxi_{j, i}(A_t)
     M_i(A_t \mid J_t)}
     {\sum^I_{k=1} \sum^I_{j=1} {\bf \varphi}_j(A^p_t) \bxi_{j, k}(A_t)
     M_i(A_t \mid J_t)}
      & \text{Otherwise},
     \end{cases}
         \label{def of varphi}
 \end{align}
  \vspace{-1.0em}
  
  \noindent
  where $A_t^p$ is $A_t$'s parent node.
 Also, let $h(J_t \mid A_t)$ be a function of $J_t$ defined as
 \vspace{-1.0em}
 \begin{align}
 h(J_t \mid A_t) =
   \sum^I_{i=1} 
   \left\{
   \sum^I_{j=1}
   {\bm \varphi}_j(A^p_t)
   \bxi_{j,i}(A_t)
   \right\}
   M_i(A_t \mid J_t) 
    \frac{ \mu(A_{t,l})^{-n(A_{t,l})}
    \mu(A_{t,r})^{-n(A_{t,r})}}
   {
   \mu(A_{t})^{-n(A_{t})}
   },
   \label{def: h in the SMC property}
 \end{align}
  \vspace{-3em}
 
 \noindent
  where $n(A)$ denotes the total number of observations included in a node $A$.
Then the target distribution $\pi_t(T_t)$ can be expressed in terms of $\pi_{t-1}(T_{t-1})$ as
\vspace{-1.0em}
 \begin{align*}
     \pi_t(T_t)  &=
     C_t
     \pi_{t-1}(T_{t-1})
     \pi_t(J_t \mid T_{t-1}) w_t(T_{t-1}),
 \end{align*}
 \vspace{-3.5em}
 
 \noindent
 where $C_t$ is a constant and 
 \vspace{-1.0em}
 \begin{align*}
\pi_t(J_t \mid T_{t-1}) &= 
\frac{ P(J_t) h(J_t \mid A_t)}{\sum_{j_t}  P(j_t) h(j_t \mid A_t)},\ 
w_t(T_{t-1}) = 
\sum_{j_t} P(j_t) h(j_t \mid A_t)
 \end{align*}
  \vspace{-3.5em}
 
 \noindent
 The summation over $j_t$ is taken over all possible decisions.
\end{prop}
 
  \begin{cor}
 \label{cor: proposal and incremental weight}
Let $h(J_t \mid A_t)$ be the function defined in Proposition~\ref{prop: components to make SMC}. 
Then the proposal distribution $\pi_t(D_t \mid T_{t-1})$ is given by
 \vspace{-1em}
 \[
    \pi_t(D_t \mid T_{t-1})
    =
    \pi_t(D_t \mid T_{t-1})
    \pi_t(L_t \mid D_t, T_{t-1}),
    \text{ where }
 \]
 \vspace{-3.5em}
 \begin{enumerate}
     \item $\pi_t(D_t \mid T_{t-1})$ is ${\rm Mult}(\tilde{\lambda}_1(A_t),\dots,\tilde{\lambda}_d(A_t))$ with
     \vspace{-1em}
  \begin{align*}
    \tilde{\lambda}_j(A_t)
    &\propto \sum^{N_L-1}_{l=1} \pi_t((j, l/N_L) \mid T_{t-1}) 
    \propto
    \lambda_j(A_t)
    \sum_{l=1}^{N_L-1} 
    \beta_l(A_t)
    h( (j, l/N_L) \mid A_t).
 \end{align*}
   \vspace{-3.5em}
 
 \noindent
 \item Given $D(A_t) = j$, the conditional posterior of $L(A_t)$ is
 \vspace{-1em}
 \begin{align*}
    \pi_t(L_t =l/N_L \mid D_t = j, T_{t-1}) = \sum^{N_L-1}_{l=1} \tilde{\beta}_l (A_t) \delta_{l/N_L} (\cdot),
 \end{align*}
  \vspace{-3.5em}
 
 \noindent
for $j=1,2,\ldots,I$ and  $l=1,\dots,N_L-1$ with
\vspace{-1em}
  \begin{align*}
    \tilde{\beta}_l(A_t)
    &\propto  \beta(A_t) h(j, l/N_L \mid T_{t-1}).
 \end{align*}
\vspace{-3.5em}
 
 \noindent
 \end{enumerate}
 We also have an analytical expression of the incremental weight:
 \vspace{-1em}
 \begin{align*}
    w_t(T_{t-1})
    &=
    \sum^d_{j=1}
    \sum_{l=1}^{N_L-1}
    \lambda_j(A_t)
    \beta_l(A_t)
    h( (j, l/N_L) \mid A_t).
 \end{align*}
\vspace{-3.5em}
 \end{cor}
 \noindent 
 \textcolor{black}{
     Remark: The recursive function ${\bf \varphi}_i(A_t)$ can be computed based on ${\bf \varphi}_i(A^p_t)$ with the fixed computational cost. Hence, the optimal proposal $\pi_t(J_t \mid T_{t-1}) $ and the incremental weight $w_t(T_{t-1})$, which are functions of $h(J_t \mid A_t)$, can be obtained at each step with constant computational cost with complexity $O(I^2 N_L d\,  n(A_t))$. As such, our inference algorithm scales linearly in both the dimensionality and the sample size. 
 }

We summarize the algorithm in updating the particle system from $\{T^m_{t-1}, W^m_{t-1}\}^M_{m=1}$ to $\{T^m_t, W^m_t\}^M_{m=1}$ below. All operations are repeated for $m = 1,\dots,M$.

\noindent
\hrulefill \vspace{-1.0em}
\begin{description}
\item[1. Choosing the current node]~\\
    From $T^m_{t-1}$, choose the \textcolor{black}{the node generated in the earliest step among the current leaf nodes that have not been terminated}, which is denoted by $A_t$. \vspace{-1.0em}
\item[2. Obtaining the information of the parent node]~\\
    Locate $A_t$'s parent node, $A^p_t$, and fetch the values of  ${\bm \varphi}_i(A^p_t)$ for $i=1,\dots,I$.\vspace{-1.0em}
\item[3. Computing the necessary quantities]~\\
    For all possible $J_t = (D_t, L_t)$, compute $M_i(A \mid J_t)$ ($i=1,\dots,I$) and $h(J_t \mid A_t)$. \vspace{-1.0em}
\item[4. Dividing the current node]~\\
    Compute the parameters $\tilde{\lambda}_j(A_t)$ for $j=1,\dots,d$ and sample 
    \vspace{-1.5em}
    \begin{align*}
    D^m_t \sim {\rm Mult}(\tilde{\lambda}_1(A_t),\dots,\tilde{\lambda}_d(A_t)).    
    \end{align*}
    \vspace{-4.0em}
    
    \noindent
    Given $D^m_t$, compute the parameters $\tilde{\beta}_l(A_t)$ for $l=1,\dots,N_L-1$ and sample
    \vspace{-1em}
    \begin{align*}
        L^m_t \sim 
        \sum^{N_L-1}_{l=1}
        \tilde{\beta}_l(A_t) \delta_{l/N_L}(\cdot).
    \end{align*}
    \vspace{-3.5em}
    
    \noindent
 Divide the current node $A_t$ with $J^m_t = (D^m_t, L^m_t)$ to update the tree $T^m_t$.  \vspace{-1.0em}
\item[5. Updating the importance weight]~\\
    Compute the incremental weight $w_t(T^m_{t-1})$ and update the importance weights
    \vspace{-1em}
    \begin{align*}
        W^m_t &=
        \frac{W^m_{t-1} w_t(T^m_{t-1})}{\sum^M_{m'=1} W^{m'}_{t-1} w_t(T^{m'}_{t-1})}.
    \end{align*}
    \vspace{-3.5em}
    
    \noindent
    If the effective sample size $1/\sum^M_{m=1} (W^m_t)^2$ is less than some prespecified threshold (e.g., $M/10$), resample the particles.  \vspace{-1.0em}
\item[6. Computing the information on the current node for its descendants]~\\
    Given $J^m_t$, compute ${\bm \varphi}_i(A_t)$ for $i=1,\dots,I$.
    \vspace{-1.5em}
\end{description}
\hrulefill

\textcolor{black}{
In the algorithm, we stop dividing $A_t$ if either the depth of $A_t$ is equal to a pre-set maximum resolution $K$ (e.g., 15) or the number of observations in $A_t$ is less than a pre-set threshold (e.g., 5). The SMC algorithm terminates when all the nodes of all the particles have been stopped.}
\textcolor{black}{
The maximum resolution $K$ controls the level of local details that the model allows to infer, and larger values of $K$ require more computational time. In a wide range of applications we have found that setting $K$ to beyond 15 to 20 leads to minimal changes in the resulting inference. 
}

A common technique in SMC is to resample the particles according to the importance weights $\{W^m_t\}^M_{m=1}$ when the effective sample size of the particles drops below a level. 
In sampling trees, however, the importance weights are affected by the choice of nodes to divide in multiple steps, and so the standard resampling scheme can be too ``short-sighted'' and often results in sacrificing promising trees prematurely. To address this issue we follow the strategy proposed in \cite{lu2013multivariate} by resampling the particles according to weights $a^m_t \propto (W^m_t)^\kappa$ for some $\kappa \in (0,1]$, and compute the new importance weights proportional to $W^m_t / a^m_t$. We generally recommend using a moderate choice of $\kappa$ such as $0.5$, which we have found satisfactory in a variety of numerical experiments, and will be our default choice in all of our later examples. 

\vspace{-1em}
\subsection{Posterior computation given sampled tree structures}
\label{subsec: Posterior computation given sampled tree structures}
\vspace{-0.5em}

The second stage of our inference strategy is to compute the posterior distributions of the latent states $V(A)$ and the conditional probabilities $\theta_g(A)$ given each sampled tree. We shall focus on \textcolor{black}{deriving a generic recipe for computing the posterior of the latent states given the tree that works for all models under consideration.} Given both the tree and the latent states, the posterior of $\theta_g(A)$ boils down to the corresponding posterior of standard PT models \textcolor{black}{on a dyadic tree, which is problem-specific as provided in the literature on each such model. }

Now suppose the SMC algorithm has produced a collection of finite trees $\{T^m\}^M_{m=1}$ along with the importance weights $\{W^m\}^M_{m=1}$. Given each tree $T^m$, it is possible to analytically calculate the exact posterior of $\{V(A)\}_{A \in T^m}$ with recursive message passing (a form of dynamic programming), which we describe below.

For $A \in \mathcal{N}(T^m)$, let $\phi_A(i)$ be the marginal likelihood on $A$ given that $V(A)=i$. \textcolor{black}{(As before we suppress the dependency of the marginal likelihood on the data to simplify notation.)} Specifically,
\vspace{-1em}
\begin{align}
    \phi_A(i) &=
    \int q(\x \mid A) P(dq \mid V(A) = i), 
    \text{ where }
    q(\x \mid A)
    = \prod^G_{g=1}
    \prod_{z \in x_{g}(A)}
    q_g(z \mid A).
    \label{def of marginal likelihood}
\end{align}
\vspace{-3em}

\noindent
In Eq.~\eqref{def of marginal likelihood}, taking the integration with respect to $P(dq \mid V(A) = i)$ is equivalent to integrating out $\theta_g(A)$ as well as the $\theta_g(A')$ and $V(A')$ terms for all descendants $A'$ of $A$.
Another useful quantity is the marginal likelihood on a node $A$ given the state of its parent node $V(A^p)=i$, which we denote as $\Phi_{A}(i)$ and is given by
\vspace{-1em}
\begin{align}
\label{eq:Phi}
    \Phi_A(i)
    &=
    \begin{cases}
         \prod_{x \in \x(A)} \mu(x \mid A) 
         & \text{if } $A$ \text{ is a leaf node},\\
         \sum_{i'=1}^{I} \bxi_{i,i'}(A) \phi_A(i')
         & \text{if } $A$ \text{ is a non-leaf node}.
    \end{cases}
\end{align}
\vspace{-2em}

\noindent
Note that the $\Phi_A(i)$ and $\phi_A(i)$ terms are related by
\vspace{-1.5em}
\begin{align}
\label{eq:phi}
    \phi_A(i) &= M_i(A \mid J(A)) \Phi_{A_l}(i) \Phi_{A_r}(i),
\end{align}
\vspace{-3.5em}

\noindent
where $M_i$ is the marginal likelihood defined in (\ref{def of MA}) given under the decision $J(A)=(D(A),L(A))$ to divide $A$ into $A_l$ and $A_r$. By iteratively computing Eqs.~\eqref{eq:Phi} and \eqref{eq:phi} in a bottom-up fashion (i.e., starting from the leaves all the way to the root), we can compute the pair $\{(\phi_A(i), \Phi_A(i)): A, i\}$ for all nodes in the tree, and these pairs are the ``messages'' passed along the tree from leaf to root. 

Given the values of $\{(\phi_A(i), \Phi_A(i)): A, i\}$, we can now obtain the posterior Markov transition probability matrices of the latent states given the tree $T$,
$\tilde{\bxi}(A) = (\tilde{\bxi}_{i,i'}(A))_{I\times I}$
where the $(i,i')$th element $\tilde{\bxi}_{i,i'}(A)=P(V(\Omega) = i' \mid V(A^p) =i ,\x, T)$ and the posterior marginal probabilities of the latent states given the tree $T$,
\vspace{-1.5em}
\begin{align*}
    \tilde{\bgamma}(A) & = (\tilde{\bgamma}_i(A))^I_{i=1} 
    = (P(V(\Omega) = i \mid \x, T))^I_{i=1}.
\end{align*}
\vspace{-3.5em}

\noindent
Specifically, by the Bayes' theorem, 
    $\tilde{\bxi}(A) = D^{-1}_1(A) \bxi(A) D_2(A),$
where $D_1(A)$ and $D_2(A)$ are diagonal matrices with $
    D_1(A)_{i,i} = \Phi_A(i)$ and
    $
    D_2(A)_{i,i} = \phi_A(i)$.
After computing these transition matrices, we can compute $\tilde{\bgamma}(A)$ (the feedback ``message'') in the top-down manner (i.e., starting from the root and down to the leaves) as $\tilde{\bgamma}(\Omega) = \tilde{\bxi}_{1,\cdot} (\Omega)$ and $\tilde{\bgamma}(A) = \tilde{\bgamma}(A^p) \tilde{\bxi}(A) \text{ for } A \neq \Omega.$

 \textcolor{black}{
  We note that $\Phi_\Omega(1)$ computed in the recursive algorithm is the overall marginal likelihood given the tree $T$, $P(\x \mid T)$, which can be used to find the {\em maximum a posteriori} (MAP) tree among the sampled trees, i.e., the sampled tree $T^{m}$ that maximizes $P(T^m \mid \x) \propto P(T^m) P(\x \mid T^m)$. We can use this ``representative'' tree, along with the conditional posterior of the latent states given this tree, to visualize and summarize the posterior inference in an interpretable way. 
 }

Now that we have completely described our inference algorithm, we provide two specific examples to demonstrate how one may use the output of the algorithm---namely the sampled trees along with the conditional posterior given the trees---to carry out inference. The first example is density estimation which involves learning the within-sample distributional features while the second is two-group comparison whose focus is on learning the between-sample structures. The inference strategies for these quintessential examples are generalizable to a variety of other tasks. 
\vspace{-0.5em}

\subsubsection*{Example 1: Density estimation}
\label{sec: density est method}
We consider the problem of estimating an unknown density based on a set of i.i.d.\ observations from that density. This corresponds to $G=1$ and so we can drop the subscript $g$ to simplify the notation.
We shall use the posterior mean density, also called the predictive density---$\E[q(\cdot) \mid \x]$---as an estimate for the density $q=dQ/d\mu$. 
\textcolor{black}{It can be computed by integrating out the random trees and the latent variables based on the SMC sample.} 
Specifically, for a given tree $T$, we define for all $A\in T$ and $i\in \{1,2,\ldots,I\}$, the following quantity
\vspace{-1.5em}
\begin{align*}
    e_A(i) := \E[Q(A) I_{\{V(A) = i\}} \mid \x].
\end{align*}
\vspace{-3.5em}

\noindent
All of these $e_A(i)$ terms can be computed together by a single top-down (i.e., root-to-leaf) recursion on the tree as given in the following proposition.
\begin{prop}
\label{prop: computing predictive density}
For the root node, $e_\Omega(i) = \tilde{\bgamma}_{1,i}(\Omega)$.
For a non-root node $A$, $e_A(i)$ can be computed recursively as
\vspace{-1em}
\begin{align*}
        e_A(i')
    &=
    \sum^I_{i=1} 
    \tilde{\bxi}_{i,i'}(A)\E[ \vartheta(A^p)\mid V(A^p) = i, T^m, \x]\, e_{A^p}(i), \text{ where} \\
\vartheta(A^p) &=
    \begin{cases}
        \theta(A^p) & \text{ if } A \text{ is the left child of $A^p$},\\
        1-\theta(A^p) & \text{ if } A \text{ is the right child of $A^p$}.
    \end{cases}
\end{align*}
\vspace{-2.5em}

\noindent
\end{prop}

Now with a recipe for obtaining all $e_i(A)$'s for a given tree, we can obtain the conditional predictive measure given the tree $T$, 
\vspace{-1.5em}
\begin{align*}
    \E[Q(B) \mid \x, T] = \sum_{A\in \mathcal{L}(T)} \frac{\mu(B \cap A)}{\mu(A)} \sum^I_{i=1} e_A(i) \quad \text{for any Borel set $B$.}
\end{align*}
\vspace{-3.5em}

\noindent
Now, given an SMC sample of $M$ trees and weights, 
 the posterior predictive density can be computed by
 \vspace{-1.5em}
 \begin{align*}
     \E[q(x) \mid \x]
     &\approx \sum^M_{m=1} W^m
    \frac{\E[Q(B^m(x)) \mid \x, T^m]}{\mu(B^m(x))},
 \end{align*}
 \vspace{-3em}

\noindent
 where $B^m(x) \in\mathcal{L}(T^m)$ the leaf node to which $x$ belongs. 
 
\subsubsection*{Example 2: Two-group comparison}
\label{subsubsection: Comparing the two hypotheses}

If we are interested in comparing two groups of observations using generalizations to the PT models described in Section \ref{sec: HMM General framework}, we shall compute the posterior probability of the two hypotheses $H_0$ and $H_1$.
For example, when $V(A)$ is defined as in Eq~\eqref{eq:2sample_state}, the posterior probability of the ``global'' null hypothesis $H_0: Q_1 = Q_2$ is given by 
\vspace{-1.5em}
\begin{align*}
    P(H_0\mid \bx) 
    &= \sum_{T \in \T}
    P(V(A) \neq 1 \text{ for all } A \in \mathcal{N}(T)  \mid T,x) P(T \mid \bx)\\
    &\approx
    \sum^M_{m=1}
    W^m
    P(V(A)\neq 1 \text{ for all } A \in \mathcal{N}(T^m)  \mid T^m, \x),
\end{align*}
\vspace{-3.5em}

\noindent
where the sum over $\T$ in the first row is over all finite trees with maximum resolution $K$ and the quantity $P(V(A) \neq 1 \text{ for all } A \in \mathcal{N}(T^m)  \mid T^m, \x)$ again is available analytically by message passing (details given in {\bf Supplementary Materials~\ref{sec: extra MRS algorithm})}.

In addition to testing the existence of any difference between groups, it is usually of interest to detect where and how the underlying distributions differ.
To this end, we can compute the ``posterior marginal alternative probability'' (PMAP) on each node $A$, along any sampled tree $T^m$:
\vspace{-1.5em}
\begin{align*}
    P(\theta_1(A) \neq \theta_2(A) \mid T^m, \x)
    =
    P(V(A) =1 \mid T^m, \x)
    =
    \tilde{\bgamma}_i(A).
\end{align*}
\vspace{-3.5em}

\noindent
Reporting the PMAPs along a representative tree such as the MAP among the sampled trees can be a particularly useful visualizing tool to help understand the nature of the underlying difference. 
One can also report on each $A$ the estimated magnitude of the difference using a notion of ``effect size'' based on the log-odds ratio \citep{soriano2017probabilistic}, $\mathrm{eff}(A) = 
     \left|
        \log \left[
            \frac{\theta_1(A)}{1 - \theta_1(A)}
        \right]
        -
        \log \left[
            \frac{\theta_2(A)}{1 - \theta_2(A)}
        \right]       
     \right|.$
In particular, one can report the posterior expected effect size $\E[\mathrm{eff}(A)\,|\, \x,T]$, which can be computed using a standard Monte Carlo (not MCMC) sample from the exact posterior given the representative tree. We will demonstrate this using a mass cytometry data set in Section~\ref{sec:mass}

\vspace{-1em}
\section{Theoretical Properties}
\label{sec: theory}
\vspace{-0.5em}

Next we investigate the theoretical properties of the proposed model. 
Previous theoretical analysis on the PT had mostly focused on establishing the marginal posterior consistency and contraction of the random measures $Q_g$ with respect to an unknown fixed truth \citep{walker2001bayesian,Costello2017polya}. We shall take a different perspective and instead investigate the asymptotic behavior of the marginal posterior of the partition tree $T$ and the latent states as these are often critical quantities of practical importance in applications. We note that once given the tree and the latent states, our model reduces to standard PTs \textcolor{black}{on a dyadic tree} and thus the posterior consistency of the random measures $Q_g$'s will follow from previous results once we establish the posterior consistency of the tree and the latent states. 
As the sample size increases, the two key theoretical questions of interest here are: 
\vspace{-0.5em}
\begin{itemize}
\item[(1)] What tree structures does the marginal posterior of $T$ concentrate around?  
\vspace{-0.5em}
\item[(2)] How does the posterior of the latent states given the tree behave?
\vspace{-0.5em}
\end{itemize}

These two questions have broad relevance in inference using PT models, and previously several authors have investigated the second question in the two-group comparison context for their variants of the PT model \citep{holmes2015two,soriano2017probabilistic}. In addressing the second question more generally, we aim to provide results that encompass these previous analyses as special cases. According to our limited knowledge, we are not aware of previous studies on the first question. 

We will address each of the two questions in turn. Throughout this section, we  consider finite PTs with maximum depth of the trees set to some value $K$. 
We use $\T^K$ to denote this collection of trees.
Also, \textcolor{black}{the asymptotic results are derived under the prior for $L(A)$ provided in Section \ref{sec: randomizing tree} which can depend on the (finite) sample size.
The case of an uniform priors on $L(A)$ independent of the sample size is included as a special case where the hyperparameter $\eta=0$.
} Finally, we consider models that satisfy {\bf Assumption~\ref{assumption: sample size and true measures}} and {\bf Assumption~\ref{assumption on state variables}} described below. The models discussed in Section~\ref{subsec: examples} all meet this requirement.\vspace{-0.5em}

\begin{assumption}
\label{assumption: sample size and true measures}
\textcolor{black}{
    For each group $g \in \{1,...,G\}$, let $n_g$ be the sample size and $P_g$ the true probability measure from which the observations are generated. We assume that 
    \vspace{-0.5em}
}
\begin{itemize}
    \item[(i)] There exists $\zeta_g \in (0,1]$ such that $ \zeta_g = \lim_{n \to \infty} \frac{n_g}{n}$ for $g \in \{1,\dots,G\}$, \textcolor{black}{where $n=n_1+\cdots+n_G$ is the total number of observations across all groups.}
    \vspace{-0.5em}
    \item[(ii)] The sampling distribution $P_g$ satisfies $P_g \ll \mu$, and
    \textcolor{black}{
         the density $p_g = dP_g / d\mu$ is positive almost everywhere.
    }
 \vspace{-0.5em}
    \end{itemize} 
Additionally, given the tree $T$ and the latent states, the parameters $\{\theta_g(A)\}^G_{g=1}$ are given one of the following priors (the model can adopt a mix of these priors for different combinations of $A$ and $V(A)$ values):
     \vspace{-0.5em}
     \begin{description}
        \item[Prior A : ] $\theta_g(A)$ independently follow a beta prior.\vspace{-0.5em}
        \item[Prior B : ] $\theta_1(A) = \cdots = \theta_G(A)$ and follow a beta prior.\vspace{-0.5em}
        \item[Prior C : ] \textcolor{black}{$\theta_1(A) = \cdots = \theta_G(A)\equiv c(A)$, some constant in $(0,1)$.}
        \vspace{-0.5em}
     \end{description}
\end{assumption}
Establishing the theoretical properties also requires a condition on the latent states. In particular, under some states, the support of the prior of the parameters $\{\theta_g(A)\}^G_{g=1}$ needs to include the true conditional probabilities.
To describe this requirement, given a tree $T \in \T^K$, let $S_i(A\mid T)$ be the support of the prior on $(\theta_1(A),\dots,\theta_G(A))$ under the state $V(A)=i$. 
Then, let $\tau(A \mid T)$ denote the collection of \textcolor{black}{``feasible states''  on $A$. (A state is ``feasible'' if the true conditional probabilities are in the support of the corresponding prior given the state.)} That is,
\vspace{-1.5em}
\begin{align*}
    \tau(A \mid T) := \{i \in \{1,\dots,I\}: 
    (P_1(A_l \mid A),\dots, P_G(A_l \mid A))
    \in 
    S_i(A \mid T)
    \}.
\end{align*}
\vspace{-3.5em}

\noindent 
The next assumption states that the prior for the latent states must give positive probability for all the latent states to all simultaneously be feasible.
\vspace{-0.5em}

\begin{assumption}
    \label{assumption on state variables}
    For every $T \in \T^K$, 
       $ P\left(V(A) \in \tau(A \mid T) \text{ for all } A\in T\right) > 0.$
\end{assumption}
\vspace{-0.5em}

\noindent
With these assumptions, we next derive asymptotic properties for the marginal posteriors for the tree and the state variables. 
In the following, we use the notation $\x_n$ instead of $\x$ for the data to indicate the total sample size.

In order to describe the posterior convergence of the partition trees, we introduce a notion for ``tree-based approximation for probability measures''. Let $T$ be a finite tree and $H$ a probability measure.
Then the ``tree-based approximation of $H$ under $T$'', denoted by $H |_T$, is  defined as
    $H |_T (B) = \sum_{A \in \mathcal{L}(T)}
     H(A) \frac{\mu(B \cap A)}{\mu(A)}.$
for any $B \in \mathcal{B}(\Omega)$.
The following theorem then characterizes the trees the posterior concentrates on as the sample size grows.
\vspace{-0.5em}
 \begin{thm}
 \label{thm: characterizign the post of trees}
 Let $\T^{K}_{M}$ be the collection of trees under which the tree-based approximation of the measures $P_g$ minimizes the Kullback-Leibler divergence from the $P_g$'s plus a penalty term on unbalanced splits. That is,
    \vspace{-1em}
\begin{align}
    \T^K_M & = \argmin_{T \in \T^K}
        \sum^G_{g=1} \zeta_g 
        \left\{
        {\rm KL}(P_g || P_g|_T)
        +
        \eta
        B_g(T)
        \right\}, \label{def of TKM}
\end{align}
\vspace{-3.5em}

\noindent
where
    \vspace{-1em}
    \[
        B_g(T)
        =
        \sum_{A \in \mathcal{N}(T)}
        P_g(A)
        f\left(
            \left|
            \frac{\mu(A_l)}{\mu(A)} - 0.5
            \right|
        \right).
    \]
    \vspace{-1.5em}
 
\noindent Then the marginal posterior of $T$ concentrates on $\T^K_M$. That is, as $n\to \infty$,
\vspace{-1em}
\begin{align*}
    P(T \in \T^K_M \mid \x_n ) \xrightarrow{p} 1.
\end{align*}
\vspace{-3.5em}

\noindent
\end{thm}

For the state variables, it is desirable that their posterior distribution concentrates on a collection of \textcolor{black}{feasible states}.
Moreover, when multiple configurations of the states are feasible, it is desirable that the posterior concentrates around such configurations that provide the most parsimonious representation of the true distributions. For example, if the true conditional distribution on a node is uniform, a model that introduces a possible non-uniform structure on this node is feasible but redundant.
\cite{white2011bayesian} and \cite{li2014bayesian} showed that, in quite general settings of multi-resolution inference, the posterior probability of such redundant models tends to concentrate its mass on 0. By adapting their techniques, we show that the same property holds in the case of our model.

To formally describe the results, we need to define the complexity of the model specified by the latent states.
Given the state $V(A) = i$, the complexity of the $\{\theta_g(A)\}^G_{g=1}$, in other words, the number of free parameters of the prior distribution under the $i$th state is denoted by $C_i(A)$. 
For example, for two-group comparison, 
\vspace{-1em}
\begin{align*}
    C_i(A)
    = 
    \begin{cases}
    1 & \text{ if } \theta_1(A) = \theta_2(A)\\
        2 & \text{ if } \theta_1(A) \neq \theta_2(A).
    \end{cases}
\end{align*}
\vspace{-2.5em}

\noindent
Next we introduce the complexity of a combination of states on the tree $T$.
Given a tree $T$, let $\bV$ denote a combination of the state variables $\{V(A)\}_{A \in \mathcal{N}(T)}$ and let $\bv = \{\bv(A)\}_{A \in \mathcal{N}(T)}$ ($\bv(A) \in \{1,\dots,I\}$) be one of the possible realizations of $\bV$.
Then we define the model complexity under $\bv$ as follows:
\vspace{-1.5em}
\begin{align}
    C(\bv) = \sum_{A \in \mathcal{N}(T)} C_{\bv(A)}(A).
    \label{def: complexity}
\end{align}
\vspace{-3.5em}

The next theorem shows that the posterior distribution of the states given the tree will concentrate on 
those that are feasible and most parsimonious.

\begin{thm}
\label{thm: posterior of state variables}
For $T \in \T^K$, let $\sV_T = \{\bv: \bv(A) \in \tau(A \mid T) \text{ for all }
A \in \mathcal{N}(T) \}$.
Then $P \left(\{\bV \in \sV_T\}
    \cap
    \left\{ C(\bV) = 
        \min_{\bv \in \sV_T}
    C(\bv)
    \right\}
    \mid T, \x_n\right) \xrightarrow{p} 1$.
\end{thm}
\noindent Remark: Consistency results for several existing models are special cases of this theorem.
For example, we derive the consistency of the MRS model for two-group comparison as a corollary in {\bf Supplementary Materials~C}. 

\vspace{-1.5em}
\section{Experiments}
\label{sec: experiments}
\vspace{-0.5em}

In this section, we carry out simulation studies to examine the performance of our model and inference algorithm. In particular, we are interested in (i) understanding how the model with the flexible tree prior compares to those with a ``divide-in-the-middle'' restriction, and (ii) verifying the linear scalability of our inference algorithm with respect to increasing dimensionality. 
We again consider the two quintessential examples---(i) density estimation and (ii) the two-group comparison---for inferring within-group and between-group structures respectively.

 We shall consider both low-dimensional settings where the underlying structure is easy to interpret and software for existing PT models are available, and high-dimensional settings for which existing implementation of PT models is not applicable and we use our SMC algorithm to carry out inference for both our model and the earlier models with fixed partitioning points (which are special cases of our model). 
 Throughout the experiments, the parameters $N_L$ and $M$ are fixed to $32$ and $1000$ respectively. We note that larger $N_L$ values can also be adopted at a linear computational cost but did not lead to noticeable change in performance in our examples.
 Details such as the settings of hyper-parameters and simulated data sets are provided in {\bf Supplementary Materials~\ref{sec: Details of the experiemnts}} unless explicitly described in this section.
 
\vspace{-1em}
\subsection{Density estimation}
\label{sec: experiments (density estimation)}
\vspace{-0.5em}
We first consider 2D examples to observe what kind of tree structures are obtained under the flexible model and how prior specification in Eq.~\eqref{prior of L} influences the performance.
After that, we move to higher dimensional cases to examine the scalability of our new SMC method and the effect of incorporating the flexible partition. For this task we compare our model with the APT model \citep{ma2017adaptive} which also incorporates a prior on the dimension to divide but restricts partitioning at middle points. Its posterior computation is implemented by the {\tt apt} function in the R package {\tt PTT}.

\vspace{-1em}
\subsubsection{Two-dimensional cases}
\label{subsubsec: 2D density estimation}
\vspace{-0.5em}

\begin{figure}[p]
\centering
\begin{tabular}{ccc}
    \includegraphics[height=4.5cm]{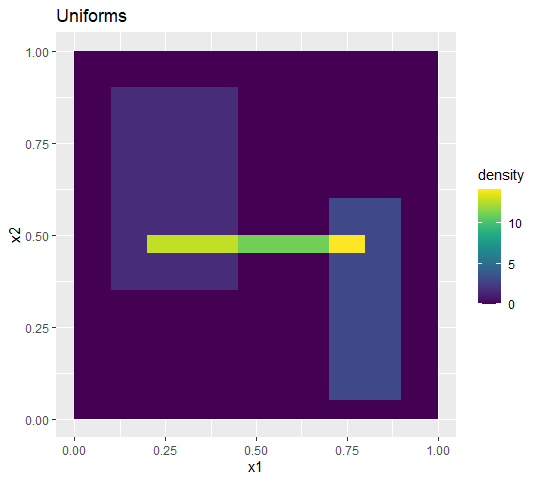}
    &
    \includegraphics[height=4.5cm]{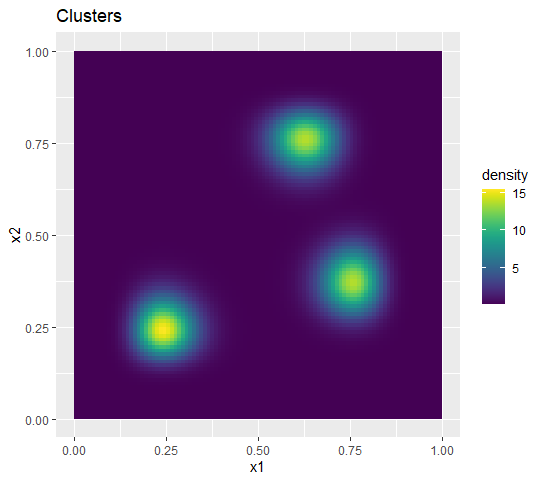}
    &
    \includegraphics[height=4.5cm]{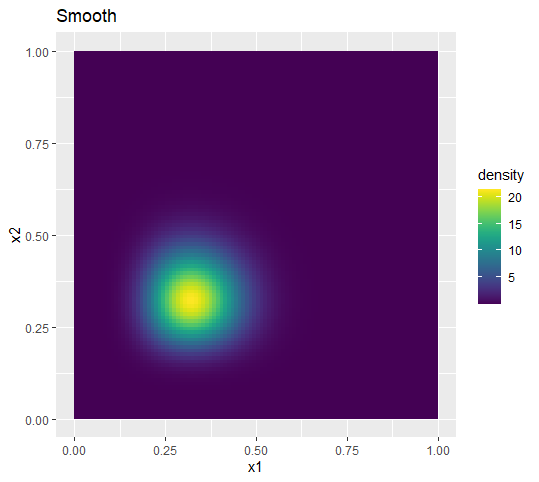}
    \\
    \includegraphics[height=4.5cm]{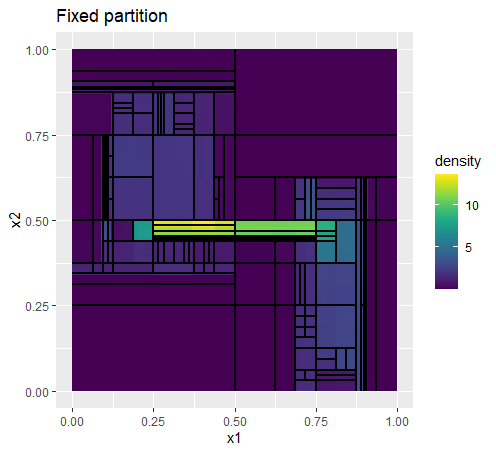}       
    &
    \includegraphics[height=4.5cm]{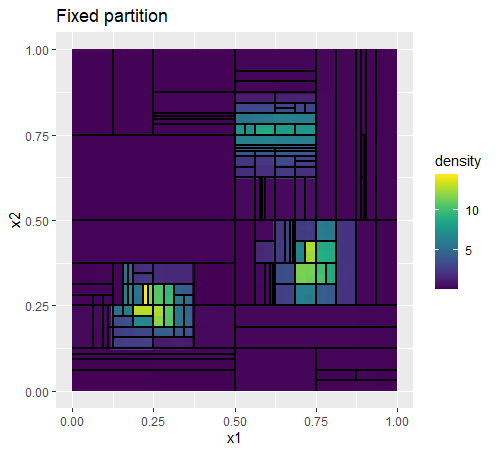}    
    &
    \includegraphics[height=4.5cm]{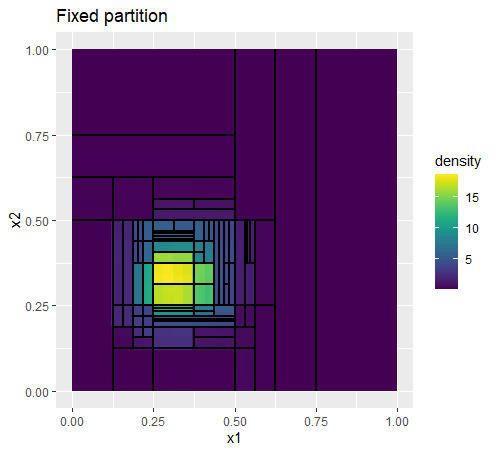}   
    \\
    \includegraphics[height=4.5cm]{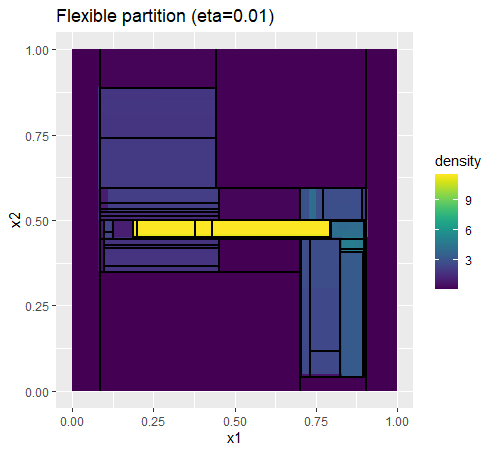}       
    &
    \includegraphics[height=4.5cm]{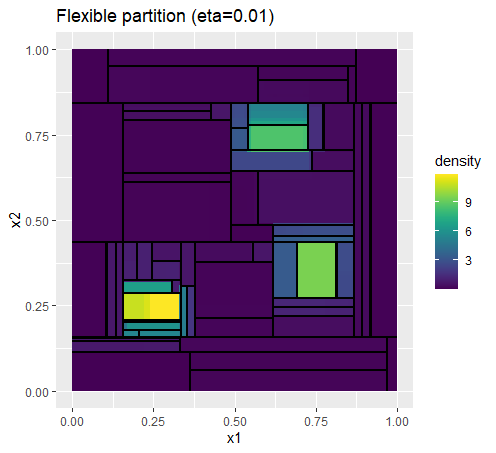}    
    &
    \includegraphics[height=4.5cm]{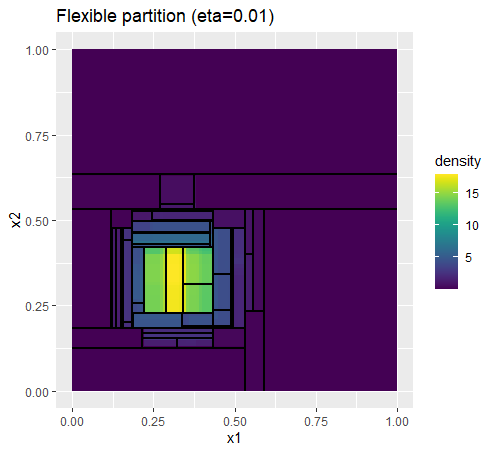}   
    \\
    \includegraphics[height=4.5cm]{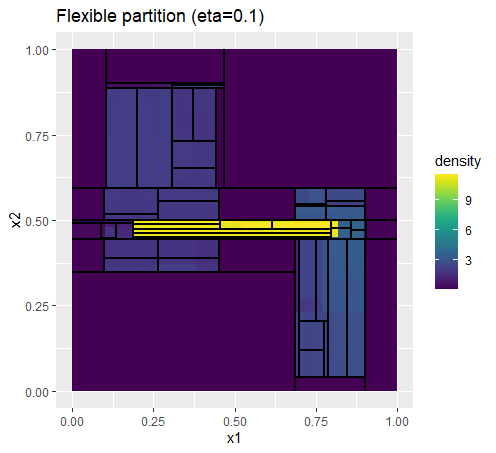}       
    &
    \includegraphics[height=4.5cm]{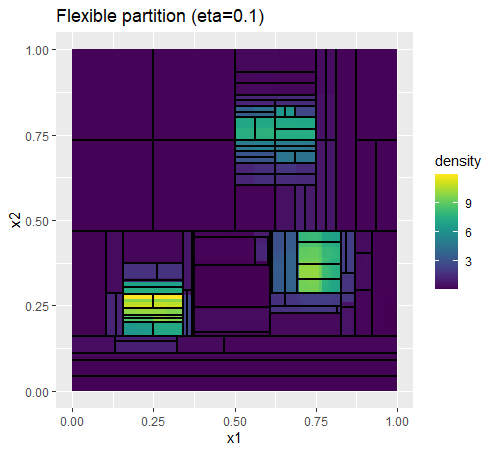}    
    &
    \includegraphics[height=4.5cm]{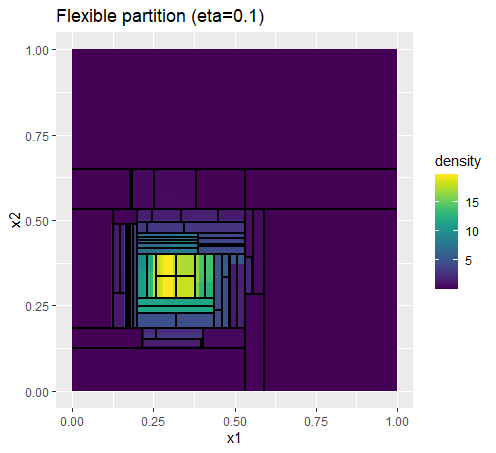}    
\end{tabular}
\caption{\textcolor{black}{The posterior means of the densities} and the representative trees obtained under $n = 1000$. Each column corresponds to a simulation scenario. The first row shows that true densities, the second row corresponds to the APT model (with fixed partition), and the third and fourth rows correspond to our model with flexible partitioning with parameters $\eta = 0.01$ and $0.1$ respectively.
}
\label{fig: estimeted densities and trees}
\end{figure}

Simulated data are generated from the three scenarios with the densities visualized in the first row of Figure \ref{fig: estimeted densities and trees}. (Details on the simulation settings are provided {\bf Supplementary Materials~\ref{sec: Details of the experiemnts}.1.2}.)
Also presented in Figure \ref{fig: estimeted densities and trees} are examples of \textcolor{black}{the posterior mean densities $ \E[q \mid \x]$ as well as the partition blocks under the MAP tree. Note that the posterior mean is computed by integrating out the unknown tree, and the MAP tree is presented to visualize key distributional features.} 
The results for the first scenario confirms that our more flexible model is much more effective in capturing the discontinuous boundaries of the true density. For the second scenario, our model tends to draw the boundaries that surround the true clusters. In the trees given under the different values of $\eta$ , however, we can see that fewer nodes were divided inside the clusters when $\eta = 0.01$.
In contrast, when $\eta = 0.1$, the representative tree draws outlines of the clusters and divides regions inside of the clusters at the same time. 
A similar phenomenon is observed in the third scenario---under our model with flexible partitioning points, partition lines are formed around the region with high density, when $\eta=0.1$ for the boundaries were also drawn within the high probability region.
The quantitative comparison based on the KL divergence is provided in {\bf Supplementary Materials~\ref{sec: additional figures}}, which is consistent with the explanation above.

\vspace{-1em}
\subsubsection{Higher-dimensional cases}
\label{subsec: two sample experiment, high-dimensional}
\vspace{-0.5em}

We generate $d$-dimensional i.i.d.\ observations from a density with independent pairs of margins, i.e., $f(x_1,x_2,\ldots,x_d)=\prod_{j=1}^{d/2} f_j(x_{2j-1},x_{2j})$ where  
\vspace{-1em}
\begin{align*}
    f_j(x_{2j-1},x_{2j})=&
    p_j {\rm Beta}(x_{2j-1}\mid 0.25, 1) \times {\rm Beta}(x_{2j}\mid 0.25, 1) \\ +
    &(1-p_j) {\rm Beta}(x_{2j-1}\mid 50/j, 50/j) \times {\rm Beta}(x_{2j}\mid 50/j, 50/j),
\end{align*}
   \vspace{-3em}
   
\noindent
with $p_j = 0.25 + 0.7/j$. 
We consider two different situations: (i) \textcolor{black}{the dimension $d=6$}, and the sample size $n$ changes from 5,000 to 50,000; and (ii) the sample size $n=10,000$ and the dimensionality changes from $10$ to $100$.
For our method, the maximum depth $K$ is set to $15$.
\textcolor{black}{
For the algorithm of the original APT (implemented by the {\tt apt} function in the {\tt PTT} package), in the first case with $d=6$, the maximum resolution is fixed to 9 because setting higher values leads to insufficient memory.
Also, because the {\tt apt} function does not scale if the dimension is beyond $d \approx 10$, in the second case with large $d$, we used the proposed SMC algorithm to carry out inference for the original APT model, which corresponds to setting $N_L = 2$, namely, fixing all the partition points to the middle. 
In addition to the original APT, we show the comparison with the classical PT method in which each node is split into $2^d$ and the maximum depth is 15, the Dirichlet process Gaussian mixture model \citep{escobar1995bayesian, muller1996bayesian} implemented by the {\tt PYdensity} function in the R package {\tt BNPmix} \citep{corradin2021bnpmix}, and for (i) the Gaussian kernel density estimation using the {\tt kde} function in the R package {\tt ks} \citep{duong2007ks}.}

\begin{figure}[thb]
\centering
\begin{tabular}{cc}
    \includegraphics[height=4.5cm]{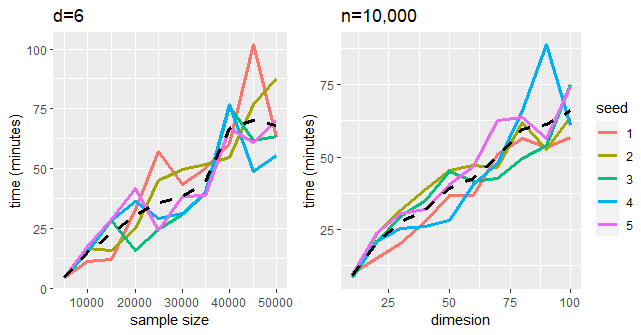}
\end{tabular}
\caption{The wall time under five different data sets. The flexible model with $\eta=0.01$ is used.
\textcolor{black}{
The black dashed lines indicate the average times.}
}
\label{fig: wall time}
\end{figure}

Figure \ref{fig: wall time} presents the computational time for five different data sets.
To obtain the result, we used a singe-core environment using \textcolor{black}{Intel Xeon Gold 6154 (3.00 GHz) CPU.}
The computational time is linear in both the sample size and the dimensionality. 

\begin{figure}[htb]
\centering
\begin{tabular}{cc}
\includegraphics[height=5.0cm]{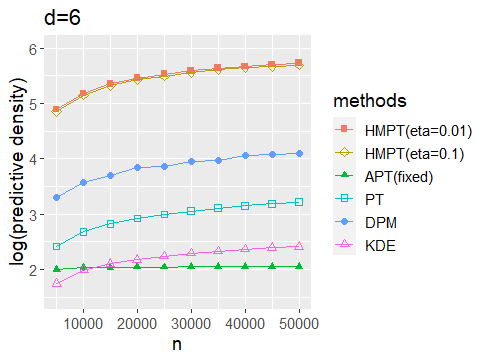}
&
\includegraphics[height=5.0cm]{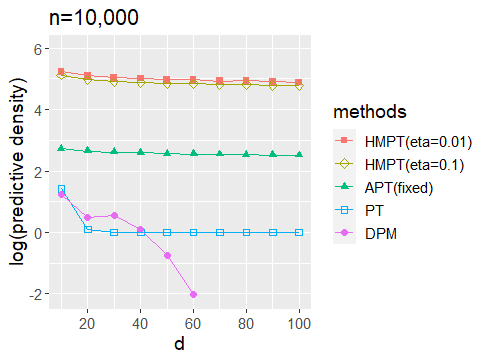}
\end{tabular}
\caption{
Predictive performance of 6 methods. Each point corresponds to the average of the predictive score in Eq.~\eqref{definition of D} based on 50 data sets. Each interval is formed by adding and subtracting the standard deviation.
In the right plot, the predictive scores of the DPM model for the over 60 dimensional cases are below the displayed range. 
}
\label{fig: multi pred}
\end{figure}

Because in the high-dimensional settings we cannot obtain the KL divergence between the estimated density and the true density through numerical approximation, we compare the models based on predictive scores, which is in essence an empirical estimate of the KL divergence. Specifically, for each training set $\bx$, we can generate a new test set denoted by $\x^* = (x^*_1, \dots, x^*_n)$ from the same true model and compute
\vspace{-1.5em}
\begin{align}
    \frac{1}{n} \sum^{n}_{i = 1} \log \hat{p}(x_i^* \mid \x),
    \label{definition of D}
\end{align}
\vspace{-3.0em}

\noindent 
where $\hat{p}$ is \textcolor{black}{the posterior mean of the density $\E[q \mid \x]$.} We repeat this computation for 50 test/training set pairs and take the average. 
The results, given in Figure \ref{fig: multi pred}, show that our model substantially outperforms the competitors by this criteria both when $d=6$ with varying sample size and when $n$ is fixed with varying dimensionality.
\textcolor{black}{
We also investigate the performance under small sample sizes, and the results are similar. (See {\bf Supplementary Materials \ref{sec: additional figures}} for details.) 
}

\vspace{-1em}
\subsection{Two-group comparison}
\vspace{-0.5em}

Next we consider the two-group comparison problem, evaluate the performance of our model, and compare it to the original MRS with the ``divide-in-the-middle'' restriction. We use three scenarios (``Local location shift'', ``Local dispersion difference'', and ``Correlation'') to generate 50-dimensional data sets. (Details of the scenarios are provided in {\bf Supplementary Materials \ref{sec: Details of the experiemnts}.2.2}.)
The first two scenarios involve two-group difference that lies in only parts of the sample space, or ``local'' differences. which will help demonstrate the usefulness of inferring the partition tree in identifying the nature of the differences. 
The sample size is $n_1 = n_2 = 2,000$ in all scenarios. 

The original algorithm for inference under the MRS model by message passing, which is implemented by the {\tt mrs} function in the {\tt R} package {\tt MRS}, is not scalable beyond about 10 dimensions even with fixed partition locations. Hence we compute the posterior for both our model and the original MRS in all scenarios with our SMC and message passing hybrid algorithm. 
\textcolor{black}{For the two models, the maximum resolution $K$ is fixed to 15.}
We compare the performance using receiver operating characteristic (ROC) curves computed based on 200 simulated data sets under each scenario.

\begin{figure}[ht]
\centering
\begin{tabular}{c}
    \includegraphics[height=5cm]{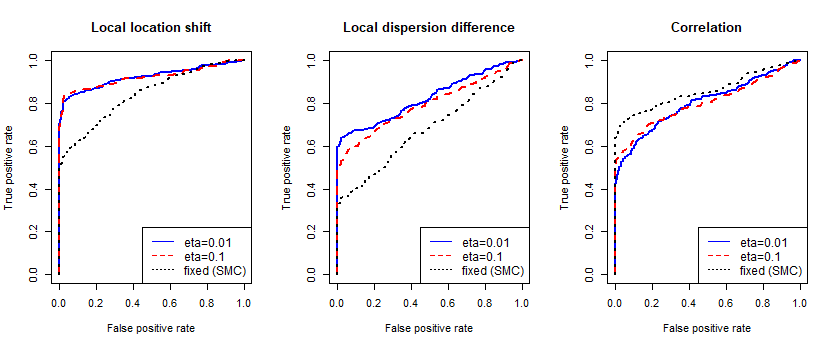}
\end{tabular}
\vspace{-0.5em}
\caption{The ROC curves for the 50-dimensional examples.}
\label{ROC}
\end{figure}

\begin{figure}[ht]
\centering
\begin{tabular}{c}
    \includegraphics[height=4.5cm]{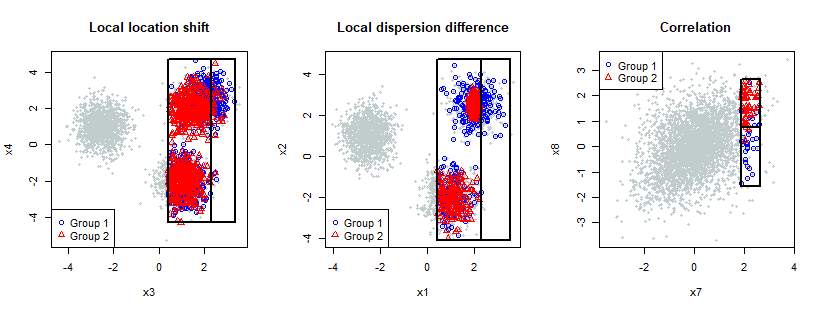}
\end{tabular}
\vspace{-0.5em}
\caption{The node with the highest PMAP $P(V(A)=1 \mid \x)$ under the three scenarios for the 50-dimensional example, estimated by the MRS with flexible partitioning and $\eta=0.1$. The solid lines mark the boundaries of the nodes and the partition line that divides them into the two children nodes. The red triangle points and the blue circle points are the observations of the two groups included in the node. Gray points indicate the observations outside the node.}
\label{fig: two_sample_problem sample_windows}
\end{figure}

Figure \ref{ROC} presents the ROC curves.
For the location shift and dispersion differences, the model with flexible partitioning results in substantially higher sensitivity. For the correlation scenario, the model with fixed partitioning locations performed slightly better. 
This is not surprising since in this scenario the difference exists smoothly over entire ranges of the dimensions without natural ``optimal'' division points, and so the performance gap is the cost for searching over more possible partition locations, none of which improves the model fit than the middle point. 
It is worth noting again that while the model with fixed partitioning performs well here, it is only with our new computational algorithm that it can be fit to data of such dimensionality.

To demonstrate the posterior model can help understand the nature of the differences, we present under each scenario the node with the highest PMAP, or $P(V(A)=1\mid \bx)=P(\theta_1(A)\neq \theta_2(A)\mid \bx)$, in Figure~\ref{fig: two_sample_problem sample_windows}.
In the location shift and dispersion difference scenarios the boundaries are away from the middle point to characterize the difference, which partly explains the sensitivity gain in adopting the flexible tree prior. 

\vspace{-1.5em}
\section{Application to a mass cytometry data set}
\label{sec:mass}
\vspace{-0.5em}
Finally, we apply our model for two-group comparison to a mass cytometry data set collected by \cite{kleinsteuber2016standardization}.
The data set records 19 different measurements including physical measurements and biomarkers on single cells in blood samples from a group of HIV patients as well as in reference samples from healthy donors.
For demonstration, we compare the sample from an individual patient sample (Patient \#1) and to that from a healthy donor to identify differences in immune cell profiles from these samples. The sample sizes are $29,226$ for the health donor and and $228,498$ for the patient, with each observation corresponding to a cell. We set $\eta = 0.1$ and the maximum depth $K$ to 25.
 
 Given the large sample sizes, the posterior probability for the global alternative $P(Q_1 \neq Q_2 \mid \x)$ is almost 1 and so is of less interest. Our focus is instead on identifying the cell subsets on which the samples differ and on quantifying such differences.
 To this end, we identify a representative tree and report the ``effect size'' (i.e., the posterior expected log-odds ratio between the two samples) on each node in a representative tree---the MAP among the sampled trees.

\begin{figure}[tb]
 \centering
\begin{tabular}{c}
    \includegraphics[height=5cm]{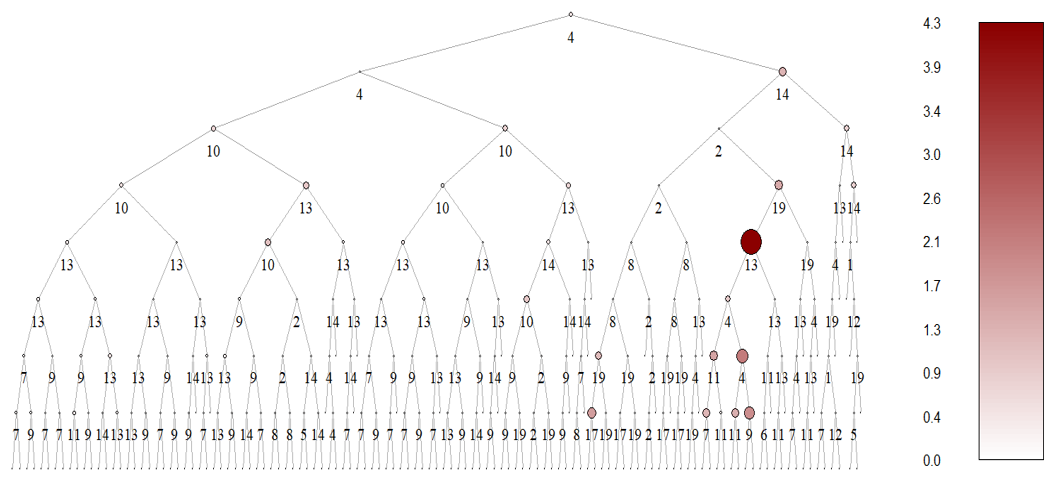}
\end{tabular}
  \caption{The MAP tree for the mass cytometry data set \textcolor{black}{visualized up to the 9th level}.  
  The size and the color indicate the estimated $\mathrm{eff}(A)$, and \textcolor{black}{the number above a node indicates dimension in which it is split.} Only the nodes with the sample size larger than 50 are drawn.}
  \label{fig: truncated tree for FC}
 \end{figure}

 \textcolor{black}{The estimated $\mathrm{eff}(A)$'s on the MAP tree up to level $9$ is visualized in Figure \ref{fig: truncated tree for FC}, and the full tree is provided in {\bf Supplementary Materials \ref{sec: additional figures}}}.
We note that the nodes on which there is significant evidence for two-group differences, as well as those with large estimated effect sizes tend to be nested or clustered in subbranches of the tree, which is consistent with our intuition that there is spatial correlation in the two-group differences, and justifies the hidden Markov structure embedded in the MRS model.

Figure \ref{fig: node with large difference (FC)} in {\bf Supplementary Materials \ref{sec: additional figures}} presents the 20 nodes with the largest values of estimated  $\mathrm{eff}(A)$. In this figure, many of the nodes are in very deep levels of the tree.
We adopted a spike-and-slab with higher spike probability in very deep tree levels to further speed up the computation (details given in {\bf Supplementary Materials \ref{sec: refixing}}) and that explains why many of the very deep, small nodes plotted have partition lines in the middle under the MAP tree. 

\vspace{-1.5em}
\section{Concluding Remarks}
\label{sec: conclusion}
\vspace{-0.5em}

We have proposed a general framework for the PT model that incorporates a flexible prior on the partition tree and can accommodate latent state variables with Markov dependency along the partition tree. We have proposed a sampling algorithm that combines SMC and recursive message passing that can scale up to moderately high-dimensional ($\sim 100$-dim) problems. Our numerical experiments confirm that our sampling algorithm scales linearly in the sample size and the flexible partitioning tree prior can result in substantial gain in performance in some settings. Though we have mainly used two inference tasks---namely density estimation and two-group comparison---to demonstrate our model and algorithm, our approach can be readily applied to other PT models with a hidden Markov structure. 

One notable limitation of our model---and in fact all CART-like models---is that we only consider trees in which the node boundaries are all parallel to the axes. This could lead to inefficiency in inference. For example, when there is a strong correlation between several variables, drawing boundaries slanted according to the correlation structure would be more effective in characterizing the underlying distribution. 
Such trees will need to be represented by more than just the $D(A)$ and $L(A)$ used in our model, and how to efficiently compute the posterior distribution is an open problem.

\textcolor{black}{Finally, our proposed algorithm is currently designed to be run in a single computer environment, so though the computational cost is linear to the sample size $n$, dealing with huge data with huge $n$ (e.g., $>10^{12}$) is not yet realistic. It is of future interest to develop parallelized versions of the algorithm for distributed systems, which could explore either the parallel structure over nodes or parallel SMC algorithms.}

\vspace{-1em}
\section*{Software}
\vspace{-0.5em}
An R package for our method is available at \url{https://github.com/MaStatLab/SMCMP}.

\vspace{-1em}
\section*{Acknowledgment}
\vspace{-0.5em}
LM's research is partly supported by NSF grants DMS-2013930 and DMS-1749789. NA is partly supported by a fellowship from the Nakajima Foundation.

\begingroup
\setstretch{1.5}
\bibliographystyle{apalike} 
\bibliography{references} 
\endgroup
 
\newpage

\appendix

\section{Spike-and-slab type prior for $L(A)$}
\label{sec: refixing}
\subsection{Introducing an auxiliary variable}
The location variable $L(A)$ follows a spike-and-slab type prior which is expressed with an auxiliary variable $R(A)$ as
\begin{align*}
    L(A) \mid R(A)
    &\sim
    \1_{\{R(A)=1\}}
    \delta_{1/2}(\cdot)
    +
    \1_{\{R(A)=0\}}
    \sum^{N_L-1}_{l=1,l/N_L\neq 1/2}
    \hat{\beta}_l(A) \delta_{l/N_L}(\cdot),
\end{align*}
where $\1$ is the indicator function and the sum of the parameters $\beta_l(A)$ is 1.
Under this prior, $L(A)$ follows the prior degenerated at 1/2 if $R(A)=1$ and otherwise follows the distribution on grid points other than the middle point.
$R(A)$ follows an asymmetric hidden Markov process
\begin{align*}
    P(R(A) = 1 \mid R(A^p) = 0)
    &= r(A), \\
    P(R(A) = 1 \mid R(A^p) = 1)
    &= 1,
\end{align*}
where $r(A) \in [0,1]$.
$R(A) = 1$ is the absorbing state, so once $A$ is divided at the middle point, $L(A') = 1/2$ for every $A$'s descendant node $A'$.  
In the estimation we especially set the parameters as follows:
\begin{align*}
    r(A) &=
    \beta_{N_L/2},\ 
    \hat{\beta}_l
    =
    \frac{\beta_l(A)}{
        \sum^{N_L-1}_{l=1,l/N_L\neq 1/2}
        \beta_l(A)
    },
\end{align*}
where $\{\beta_l(A)\}_{l=1,\dots,N_L}$ is given in (\ref{prior of L}).
Under this setting the prior of $L(A)$ satisfies 
\begin{align*}
    P(L(A) =l \mid R(A^p) = 0)
    &= \beta_l(A),\ l=1,\dots,N_L-1.
\end{align*}
Hence, $L(A)$ follows the same prior as defined in (\ref{prior of d and l}) unless $A$'s parent node is divided at the middle point, so the spike-and-slab prior can be seen as a natural extension.

\subsection{SMC algorithm}
In the SMC algorithm, we sample values of $R(A)$ in addition to $D(A)$ and $L(A)$. If $R(A)=1$, which is equivalent to $L(A)=1/2$, is sampled, we conclude there is no interesting structure on the node $A$ so fix $L$ to 1/2 for all the subsequent nodes. 
Hence, we need to generalize the SMC algorithm discussed in Section \ref{subsec: smc algorithm} to sample from the joint posterior distribution of the finite trees and the auxiliary variables $R$.

To describe this joint posterior, let $T_t$ denote the finite tree structure, which is determined by the sequence of decisions $J_{1:t}$ dividing the nodes $A_{1:t}$, and let $R_{1:t}$ be a sequence of the re-fixing variables for $A_{1:t}$.
Then the target distribution we want to sample from in the SMC is defined as
\begin{align*}
    \pi_t(T_t, R_{1:t})
    &= P(T_t, R_{1:t} \mid \x) \propto P(T_t, R_{1:t}) P(\x \mid T_t, R_{1:t}).
\end{align*}
The prior $P(T_t, R_{1:t}) = P(J_{1:t}, R_{1:t})$ have a Markov chain structure on the tree, and its transition probability is decomposed as
\begin{align*}
    P(J_t, R_t \mid J^p_t, R^p_t)
    &= P(R_t \mid R^p_t) P(J_t).  
\end{align*}
where $J^p_t = J(A^p_t)$ and $R^p_t = R(A^p_t)$ ($A^p_t$ is the parent node of $A_t$).
On the other hand, because $R_{1:t}$ are conditionally independent of the observations given $T_t$, the likelihood only depends on $T_t$ as follows:
\begin{align*}
    P(\x \mid T_t, R_{1:t})
    =
    P(\x \mid T_t).
\end{align*}
The likelihood has the same form as in the original case without the auxiliary variables $R$'s.
Thus we can obtain the following proposition as a generalization of Proposition \ref{prop: components to make SMC}.
 \begin{prop}
 \label{prop: SMC}
 Let $h(J_t \mid A_t)$ be a function of $J_t$ defined as
 \begin{align*}
 h(J_t \mid A_t) =
   \sum^I_{i=1} 
   \left\{
   \sum^I_{j=1}
   {\bm \varphi}_j(A^p_t)
   \bxi_{j,i}(A_t)
   \right\}
   M_i(A_t \mid J_t) 
   \frac{\mu(\x(A_{t,l}) \mid A_{t,l})\mu(\x(A_{t,r}) \mid A_{t,r})}
   {
   \mu(\x(A_t) \mid A_t)
   }.
 \end{align*}
Then the target distribution $\pi_t(T_t, R_{1:t})$ is expressed with $\pi_t(T_{t-1}, R_{1:t-1})$ as
 \begin{align*}
     \pi_t(T_t, R_t)  &=
     C
     \pi_t(T_{t-1}, R_{1:t})
     \pi_t(R_t \mid T_{t-1}, R^p_t)
    \pi_t(J_t \mid T_{t-1}, R_t)
     w_t(T_{t-1}, R^p_t),
 \end{align*}
 where $C$ is a constant and 
 \begin{align*}
 \pi_t(R_t \mid T_{t-1},  R^p_t)
 &=
 \frac{P(R_t \mid R^p_t) 
    \sum_{j_t} P(j_t \mid R_t) h(j_t \mid A_t)
 }{
 \sum_{i=0,1}
 P(R_t = i\mid R^p_t) 
 \left\{
 \sum_{j_t} P(j_t \mid R_t=i) h(j_t \mid A_t)
 \right\}
 }
 , \\
\pi_t(J_t \mid T_{t-1}, R_t) &= 
\frac{ P(J_t \mid R_t) h(J_t \mid A_t)}{\sum_{j_t}  P(j_t \mid R_t) h(j_t \mid A_t)},\\
w_t(T_{t-1}, R^p_t) &= 
 \sum_{i=0,1}
 P(R_t = i\mid R^p_t) 
  \left\{
 \sum_{j_t} P(j_t \mid R_t=i) h(j_t \mid A_t)
 \right\}.
 \end{align*}
 The summation with $j_t$ is taken over all possible decisions. 
 \end{prop}
 Its proof is essentially the same as Proposition \ref{prop: components to make SMC} and Corollary \ref{cor: proposal and incremental weight} so it is omitted in this material.
The conditional posteriors $\pi_t(R_t \mid R^p_t, T_{t-1})$ and $\pi_t( J_t \mid R_t, T_{t-1})$ are analytically obtained as follows.
 First, if $R^p_t = 0$, $\pi_t(R_t \mid R^p_t, T_{t-1})$ is $Bernoulli(\tilde{r}(A_t))$, where 
 \begin{align*}
    \tilde{r}(A_t) &=
     r(A_t)
    \sum^J_{j=1}
    \lambda_j(A_t)
    h((j, 1/2) \mid A_t)
    \\
    &
    \times
    \left[
     r(A_t)
    \sum^J_{j=1}
    \lambda_j(A_t)
    h((j, 1/2) \mid A_t)
    +
    (1-r(A_t))
    \sum^J_{j=1}
    \sum^{N_L-1}_{l=1}  
    \lambda_j(A_t)
    \beta_l(A_t)
    h((j, l/N_L) \mid A_t)
    \right]^{-1},
 \end{align*}
If $R^p_t = 1$, then $R_t$ is fixed to $1$.
Second, if $R_t = 0$, the posterior of $D_t$ and $L_t$ is the same distribution given in Section \ref{subsec: smc algorithm}.
 On the other hand, if $R_t = 1$, $L_t$ is fixed to $1/2$, and the conditional posterior $\pi_t(J_t \mid L_t,R_t,T_{t-1}) $ is Mult$(\hat{\lambda}_1(A_t), \dots, \hat{\lambda}_d(A_t))$, where
 \begin{align*}
     \hat{\lambda}_j(A_t)
     &
     \propto
     \lambda_j(A_t)
     h(j, 1/2 \mid A_t).
 \end{align*}
After sampling $(R_t,J_t)$, the incremental weight $w_t(T_{t-1}, R^p_t)$ ($R^p_t = 0,1$) is computed as
 \begin{align*}
     &w_t(T_{t-1}, 0) \\
     &=
     r(A_t)
    \sum^J_{j=1}
    \lambda_j(A_t)
    h((j, 1/2) \mid A_t)
    +
    (1-r(A_t))
    \sum^J_{j=1}
    \sum^{N_L-1}_{l=1}  
    \lambda_j(A_t)
    \beta_l(A_t)
    h((j, l/N_L) \mid A_t), \\
   &w_t(T_{t-1}, 1) 
   =
       \sum^J_{j=1}
    \lambda_j(A_t)
    h((j, 1/2) \mid A_t),
 \end{align*}
 with which we update the importance weight $W_t$ as $W_t \propto W_{t-1} w_t(T_{t-1}, R^p_t)$.
 \\
 
 The procedure to update the particle system $\{T^m_{t-1}, W^m_{t-1}\}^M_{m=1}$ to obtain $\{T^m_t, W^m_t\}^M_{m=1}$ is described in the following algorithm. 
The operations involving the index $m$ is repeated for $m = 1,\dots,M$.

\noindent
\hrulefill
\begin{description}
\item[1. Choosing the current node]~\\
    From $T^m_{t-1}$, choose the oldest note from the leaf nodes, which is denoted by $A_t$. 
\item[2. Obtaining the information of the parent node]~\\
    Pick up $A_t$'s parent node, which is denoted by $A^p_t$, and load the values of  ${\bm \varphi}_i(A^p_t)$ for $i=1,\dots,I$ and $R^{m,p}_t = R(A^p_t)$.
\item[3. Computing the necessary quantities]~\\
If $R^{m,p}_t=0$, compute $M_i(A \mid j,l/N_L)$ ($i=1,\dots,I$) and $h(j,l/N_L \mid A_t)$ for $j=1,\dots,d$ and $l=1,\dots,N_L-1$. \\
If $R^{m,p}_t=1$, compute $M_i(A \mid j,1/2)$ ($i=1,\dots,I$) and $h(j,1/2 \mid A_t)$ for $j=1,\dots,d$.
\item[4. Deciding whether to fix the partition or not]~\\
If $R^{m,p}_t=0$, compute the parameter $\tilde{r}(A_t)$ and draw $R^m_t \sim Bernoulli(\tilde{r}(A_t))$.\\
If $R^{m,p}_t=1$, set $R^{m}_t$ to be 1.
\item[5. Dividing the current node]~\\
Sample $J^m_t = (D^m_t, L^m_t)$ as follows:
\begin{itemize}
    \item  If $R^{m}_t=0$, compute the parameters $\tilde{\lambda}_j(A_t)$ for $j=1,\dots,d$ and sample 
\begin{align*}
D^m_t \sim {\rm Mult}(\tilde{\lambda}_1(A_t),\dots,\tilde{\lambda}_d(A_t)).    
\end{align*}
Given $D^m_t$, compute the parameters $\tilde{\beta}_l(A_t)$ for $l=1,\dots,N_L-1$ and sample
\begin{align*}
    L^m_t \sim 
    \sum^{N_L-1}_{l=1}
    \tilde{\beta}_l(A_t) \delta_{l/N_L}(\cdot).
\end{align*}
\item If $R^{m}_t=1$, compute the parameters $\hat{\lambda}_j(A_t)$ for $j=1,\dots,d$ and sample 
\begin{align*}
D^m_t \sim {\rm Mult}(\hat{\lambda}_1(A_t),\dots,\hat{\lambda}_d(A_t)),
\end{align*}
and set $L^m_t = 1/2$.
\end{itemize}
 Divide the current node $A_t$ with $J^m_t = (D^m_t, L^m_t)$ to update the tree $T^m_t$. 
\item[6. Storing the information of the state's posterior]~\\
    Given $J^m_t$, compute ${\bm \varphi}_i(A_t)$ for $i=1,\dots,I$ and store them to the memory.
\item[7. Updating the importance weight]~\\
    Compute the incremental weight $w_t(T^m_{t-1}, R^{m,p}_t)$ and update the importance weights as
    \begin{align*}
        W^m_t &=
        \frac{W^m_{t-1} w_t(T^m_{t-1}, R^{m,p}_t)}{\sum^M_{m'=1} W^{m'}_{t-1} w_t(T^{m'}_{t-1}, R^{m',p}_t)}.
    \end{align*}
    If the effective sample size $1/\sum^M_{m=1} (W^m_t)^2$ is less than some prespecified threshold (M/10, say), resample the particles. 
\end{description}
\hrulefill

\section{Additional algorithm for the MRS model}
\label{sec: extra MRS algorithm}
To describe the algorithm proposed in \cite{soriano2017probabilistic}, we keep using the same notations in Section \ref{subsubsection: Comparing the two hypotheses}.
Given the tree structure $T$, we compute functions $\tilde{\psi}(A)$ for $A \in \mathcal{N}(T)$ in the bottom-up (from the leaf nodes to the root node) manner as follows:
\begin{align*}
    \tilde{\psi}(A) = 
    \begin{cases}
        \tilde{\bxi}_{2,2} (A) + \tilde{\bxi}_{2,3} (A)
        &
        \text{ if }
        A \in \mathcal{L}(T),  \\
        \tilde{\bxi}_{2,2} (A)\tilde{\psi}(A_l) \tilde{\psi}(A_r)
         +  
        \tilde{\bxi}_{2,3} (A)
                &
        \text{ if } A \in \mathcal{N}(T)
        \setminus \{ \Omega \}
        ,\\
        \tilde{\bxi}_{1,2} (A)\tilde{\psi}(A_l) \tilde{\psi}(A_r)
        +  
        \tilde{\bxi}_{1,3} (A)
                &
        \text{ if } A = \Omega.
    \end{cases}
\end{align*}
Recall that only the first row of $\bxi(\Omega)$ is meaningful as the initial distribution. Then we obtain $\tilde{\psi}(\Omega) = P(H_0 \mid T, \x)$.

\section{Proofs}
\subsection{Posterior computation}

\begin{lem} 
\label{lem: conditinal post of the state given the tree}
    For the finite tree $T_t$, let $A_s$ be a node whose children nodes are leaf nodes. Then we have 
    \begin{align*}
        \pi_t(V_s = i \mid T_t) = \frac{\pi_t(T_t, V_s = i)}{\pi_t(T_t)}
        = 
        {\bm \varphi}_i(A_s),
    \end{align*}
    where ${\bm \varphi}_i(A_s)$ is defined in (\ref{def of varphi}).
\end{lem}
\noindent
(Proof)
Suppose that $A_s$ belongs to the $k$th layer of $T_t$.
Then there is a sub-sequence $\{\rho(l)\}^k_{l=1}$ such that $A_{\rho(l)}$ belongs to the $l$th layer and 
\begin{align*}
    \Omega = A_{\rho(1)} \supset A_{\rho(2)} \supset \cdots \supset
     A_{\rho(k)} = A_s.
\end{align*}
By the definition of $\pi_t(T_t, V_{1:t})$, for a sequence $\{v_l\}^k_{l=1}$ such that $v_l \in \{1,\dots,I\}$, we obtain the expression of the conditional posterior of $\{V_{\rho(l)}\}^k_{l=1}$ as 
\begin{align*}
    \pi_t(\{V_{\rho(l)}\}^k_{l=1} = \{v_l\}^k_{l=1} \mid T_t)
    &\propto 
    P(\{V_{\rho(l)}\}^k_{l=1} = \{v_l\}^k_{l=1})
    \prod^k_{l=1}
    M_{v_l} (A_{\rho(l)} \mid J_{\rho(l)}) \\
    &=
    \prod^k_{l=1} 
    \bxi_{v_{l-1}, v_l} (A_{\rho(l)}) M_{v_l} (A_{\rho(l)} \mid J_{\rho(l)}),
\end{align*}
where $v_0 = 1$.
We show that for every $k = 1,\dots,K$ 
\begin{align}
     \pi_t(V_{\rho(k)} = v_k \mid T_t)
     &\propto
     \sum^I_{v_1=1} \cdots \sum^I_{v_{k-1}=1}
    \left\{
    \prod^{k}_{l=1} 
    \bxi_{v_{l-1}, v_l} (A_{\rho(l)}) M_{v_l} (A_{\rho(l)} \mid J_{\rho(l)})
    \right\} 
    \notag
    \\
    &\propto {\bm \varphi}_{v_{k}}(A_s) 
    \label{conclusion of induction}
\end{align}
holds by induction.
First, if $k = 1$, which is equivalent to $s=1$, $\rho(1) = 1$, and  $A_s = \Omega$, the posterior of $V(\Omega)$ is written as
\begin{align*}
\pi_t(V(\Omega) = v_1 \mid T_t)
& \propto \bxi_{1, v_1}(\Omega) M_{v_1}(A_1 \mid J_1)
\propto
{\bm \varphi}_{v_1}(\Omega). 
\end{align*}
Second, assume that (\ref{conclusion of induction}) holds for $k= \bar{k}$.
Then, if $k = \bar{k}+1$, we have 
\begin{align*}
    &\pi_t(V_{\bar{k}} = v_{\bar{k}}, V_{\bar{k}+1} = v_{\bar{k}+1} \mid T_t) \\
        &\propto
    \sum^I_{v_1=1} \cdots \sum^I_{v_{\bar{k}-1}=1}
    \pi_t(\{V_{\rho(l)}\}^{\bar{k}+1}_{l=1} = \{v_l\}^{\bar{k}+1}_{l=1} \mid T_t)\\
    &\propto
    \sum^I_{v_1=1} \cdots \sum^I_{v_{\bar{k}-1}=1}
    \left\{
    \prod^{\bar{k}}_{l=1} 
    \bxi_{v_{l-1}, v_l} (A_{\rho(l)}) M_{v_l} (A_{\rho(l)} \mid J_{\rho(l)})
    \right\}
    \bxi_{v_{\bar{k}}, v_{\bar{k}+1}} (A_s) M_{v_{\bar{k}+1}} (A_{s} \mid J_{s})\\
    &\propto  {\bm \varphi}_{v_{\bar{k}}}(A_{\rho(\bar{k})}) \bxi_{v_{\bar{k}}, v_{\bar{k}+1}}  (A_s) M_{v_{\bar{k}+1}} (A_{s} \mid J_{s}),
\end{align*}
from which we obtain
\begin{align*}
    \pi_t(V_{\bar{k}+1} = v_{\bar{k}+1} \mid T_t) &=
    \sum^I_{v_{\bar{k}}=1} \pi_t(V_{\bar{k}} = v_{\bar{k}}, V_{\bar{k}+1} = v_{\bar{k}+1} \mid T_t)\\
    &\propto 
    \sum^I_{v_{\bar{k}}=1} {\bm \varphi}_{v_{\bar{k}}}(A_{\rho(\bar{k})}) \bxi_{v_{\bar{k}}, v_{\bar{k}+1}}  (A_s) M_{v_{\bar{k}+1}} (A_{s} \mid J_{s})\\
    &\propto {\bm \varphi}_{v_{\bar{k}+1}} (A_s). \qed 
\end{align*}

\subsubsection*{Proof of Proposition \ref{prop: components to make SMC}}
Let the finite tree $T_t$ consist of a sequence of decisions $J_{1:t} = \{J_s\}^t_{s=1}$, which sequentially divides nodes $A_{1:t} = \{A_s\}^t_{s=1}$.
To derive the proposition for the marginal posterior of $T_{t}$, we first consider the joint posterior of $T_t$ and a sequence of the state variables $V_{1:t}$ which are defined for the nodes $A_{1:t}$. 
From the structure of the model, the joint posterior is written as
\begin{align}
    \pi_t(T_t, V_{1:t}) &= P(J_{1:t}, V_{1:t} \mid \x) \notag \\
    &=
    \frac{1}{Z_t}
    P(J_{1:t})P(V_{1:t}) \prod^t_{s=1, A_s \in \mathcal{N}(T_t)}
    M_{V_s}(A_s \mid J_s)
    \prod^t_{s=1, A_s \in \mathcal{L}(T_t)} \mu(\x(A_s) \mid A_s), 
   \label{def of joint post}
\end{align}
where $Z_t$ is the normalizing constant, and $A_{s,l}$ and $A_{s,r}$ are the children nodes of $A_s$.
\\

For $T_{t-1}$ and $T_t$, since $A_t$ is divided into $A_{t,l}$ and $A_{t,r}$, we have
\begin{align*}
    \mathcal{N}(T_t) &= 
    \mathcal{N}(T_{t-1}) \cup \{A_t\}, \\
    \mathcal{L}(T_t) &= 
    \mathcal{L}(T_{t-1}) \setminus \{A_t\}
    \cup
    \{A_{t,l}, A_{t,r}\}.
\end{align*}
With the expression of the joint posterior in (\ref{def of joint post}), we obtain
\begin{align}
    \pi_t(T_t, V_{1:t})
    &=
    \frac{Z_t}{Z_{t-1}}
    \pi_{t-1} (T_{t-1}, V_{1:t-1})
    P(J_t) P(V_t \mid V_{1:t-1}) 
    M_{V_t}(A_t \mid J_t) 
    \frac{\mu(\x(A_{s,l}) \mid A_{t,l})\mu(\x(A_{s,r}) \mid A_{s,r})}
   {
   \mu(\x(A_s) \mid A_s)
   }.
   \label{relationship between pi_t and pi_t-1}
\end{align}
Let $A^p_t$ denote the parent node of $A_t$ and $V^p_t = V(A^p_t)$. Then, since the state variables follow the hidden Markov process, $P(V_t \mid V_{1:t-1}) = \bxi_{V^p_t, V_t}(A_t)$.
Integrating out $V_{1:t-1} \setminus V^p_t$ in (\ref{relationship between pi_t and pi_t-1}) gives
\begin{align*}
    \pi_t(T_t, V^p_t, V_t)
    &=
    \frac{Z_t}{Z_{t-1}}
    \pi_{t-1} (T_{t-1}, V^p_t)
    P(J_t)
    \bxi_{V^p_t, V_t}(A_t) M_{V_t}(A_t \mid J_t) 
    \frac{\mu(\x(A_{s,l}) \mid A_{t,l})\mu(\x(A_{s,r}) \mid A_{s,r})}
   {
   \mu(\x(A_s) \mid A_s)
   }.
\end{align*}
Because $A_t$ is a leaf node of $T_{t-1}$, by Lemma \ref{lem: conditinal post of the state given the tree},
we have 
\begin{align*}
    \pi_{t-1} (T_{t-1}, V^p_t=j)
    &= 
    \pi_{t-1} (T_{t-1}) \pi_{t-1}( V^p_t=j \mid T_{t-1})\\
    &=
    \pi_{t-1} (T_{t-1}) {\bm \varphi}_{j}(A^p_t).
\end{align*}
Hence, we obtain the expression of the marginal distribution of $T_t$ as
\begin{align*}
    \pi_t(T_t)
    &=
    \sum^I_{i=1} \sum^I_{j=1} \pi_t(T_t, V^p_t=j, V_t=i) \\
    &=
    \frac{Z_t}{Z_{t-1}}\pi_{t-1} (T_{t-1})
    P(J_t)
    \sum^I_{i=1} 
   \left\{
   \sum^I_{j=1}
   {\bm \varphi}_j(A^p_t)
   \bxi_{j,i}(A_t)
   \right\}
   M_i(A_t \mid J_t) 
   \frac{\mu(\x(A_{t,l}) \mid A_{t,l})\mu(\x(A_{t,r}) \mid A_{t,r})}
   {
   \mu(\x(A_t) \mid A_t)
   },
\end{align*}
which completes the proof. \qed

\subsubsection*{Proof of Corollary \ref{cor: proposal and incremental weight}}
\textcolor{black}{
    For $\pi_t(D_t \mid T_{t-1})$, by Proposition \ref{prop: components to make SMC}, we obtain 
    \begin{align*}
    \pi_t(D_t = j \mid T_{t-1})
    &=
    \sum^{N_L-1}_{l=1} \pi_t((j, l/N_L) \mid T_{t-1}) \\
    &\propto
    \sum_{l=1}^{N_L-1} 
    P(D_t = j, L_t = l/N_L)
    h( (j, l/N_L) \mid A_t)\\
    &\propto
    \lambda_j(A_t)
    \sum_{l=1}^{N_L-1} 
    \beta_l(A_t)
    h( (j, l/N_L) \mid A_t).
    \end{align*}
    On the other hand, the conditional probability of $L_t$ is obtained as follows:
    \begin{align*}
        \pi_t(L_t = l/N_L \mid D_t = j, T_{t-1})
        &\propto 
        P(D_t = j, L_t = l/N_L)
    h( (j, l/N_L) \mid A_t)\\
    &\propto 
    \beta(A_t) h(j, l/N_L \mid T_{t-1}).
    \end{align*}
    The expression of $w_t(T_{t-1})$ immediately follows Proposition \ref{prop: components to make SMC}.
    \qed
}

\subsubsection*{Proof of Proposition \ref{prop: computing predictive density}}
In this discussion, we suppress $\x$ and $T^m$ in the expectation for simplicity. 
First, when $A=\Omega$, by the definition $e_\Omega(i) = \tilde{\bgamma}_{1,i}(\Omega)$.
Next, if $A$ is not the root node, we can decompose $e_A(i')$ as
\begin{align*}
    e_A(i')
    &=
    \sum^I_{i=1} 
    \E[ Q(A) I[V(A) = i']I[V(A^p) = i]].
\end{align*}
For the summand, because $\theta(A^p)$ and $V(A)$ are conditionally independent given $V(A^p)$, we obtain
\begin{align*}
    &\E[ Q(A) I[V(A) = i']I[V(A^p) = i]] \\
    &=
    \E[\E[ \vartheta(A^p) I[V(A) = i']\mid V(A^p)]I[V(A^p) = i]  Q(A^p)] \\
    &=
    \E[\E[ \vartheta(A^p) I[V(A) = i']\mid V(A^p) = i]I[V(A^p) = i]  Q(A^p)] \\
    &=\E[\E[ \vartheta(A^p)\mid V(A^p) = i] P(V(A) = i'\mid V(A^p) = i) I[V(A^p) = i]  Q(A^p)]\\
    &=
    \tilde{\bxi}_{i,i'}\E[ \vartheta(A^p)\mid V(A^p) = i] e_{A^p}(i).
\end{align*}
Therefore, we obtain
\begin{align*}
    e_A(i')
    &=
    \sum^I_{i=1} 
    \tilde{\bxi}_{i,i'}(A)\E[ \vartheta(A^p)\mid V(A^p) = i] e_{A^p}(i).
\end{align*}

\subsection{Asymptotic properties of the tree posteriors}

\begin{lem}
\label{lem: limit of log of sum}
\textcolor{black}{
    Let $\{X^i_n\}_{n=1,2,\dots}$ ($i=1,\dots,L$) be sequences of random variables that satisfy the following conditions:
    \begin{enumerate}
        \item $X^i_n > 0$ for every $i$ and $n$.
        \item As $n \to \infty$, the following convergence occurs
        \[
            \frac{\log X^i_n }{n}
            \xrightarrow{p} c_i,
        \]
        where $c_1 \geq c_2 \geq \dots \geq c_L$.
    \end{enumerate}
    Then we have 
    \[
        \frac{\log \sum^L_{i=1}  X^i_n}{n} \xrightarrow{p} c_1.
    \]
    }
\end{lem}
\noindent
\textcolor{black}{
(Proof)
We only discuss the case where there exists $l$ such that $c_l > c_{l+1}$.
Let $\epsilon > 0$.
Then, by the first condition,
\begin{align*}
    P
    \left(
    \frac{\log \sum^L_{i=1} X^i_n}{n} < c_1 - \epsilon
    \right)
    \leq 
    P
    \left(
    \frac{\log X^1_n}{n} < c_1 - \epsilon
    \right)
    \to 0
\end{align*}
as $n \to \infty$.
On the other hand, 
\begin{align*}
    &P
    \left(
    \frac{\log \sum^L_{i=1} X^i_n}{n} < c_1 + \epsilon
    \right) \\
    &\geq
    P
    \left(
    \left\{
    \frac{\log L \max_{i=1,\dots,L} X^i_n}{n} < c_1 + \epsilon
    \right\}
    \cap
    \bigcap^l_{i=1}   \bigcap^L_{j=l+1}
    \{ X^i_n > X^j_n \}
    \right)\\
    &\geq
    P
    \left(
    \left\{
    \frac{\log L \max_{i=1,\dots,l} X^i_n}{n} < c_1 + \epsilon
    \right\}
    \cap
    \bigcap^l_{i=1}   \bigcap^L_{j=l+1}
    \{ X^i_n > X^j_n \}
    \right) \to 1.
    \qed
\end{align*}
}

\begin{lem}
\label{lem: asymptotic behavior for a single node}
        For $T \in \T^K$ and $A \in \mathcal{N}(T)$, if $i \in \tau(A \mid T)$, then
\begin{align*}
    &\frac{\log M_i(A \mid j_A)}{n}
    \xrightarrow{p}
    \sum^G_{g=1}
    \zeta_g P_g(A)
    \left[
        P_g(A_l \mid A) \log P_g(A_l \mid A)
        +
        P_g(A_r \mid A) \log P_g(A_r \mid A)
    \right],
\end{align*}
where $j_A$ is the splitting rule that divides $A$ into $A_l$ and $A_r$.
\end{lem}
\noindent
(Proof)
By the result of \cite{schwarz1978estimating}, since the parameter $\theta(A)$ follow the beta distribution, which belongs to a continuous exponential family, the log of the marginal likelihood is written as
\begin{align}
    \log M_i(A\mid T)
    &= 
    \hat{l}_A(i,T) - \frac{r_i}{2} \log n(A) + \mathcal{O}_p(1), \notag \\
    \hat{l}_A(i,T)
    &=
    \log
    \left[
        \prod^G_{g=1}
        \hat{\theta}_g(A)^{n_g(A_l)}
        (1 - \hat{\theta}_g(A))^{n_g(A_r)}
    \right] \notag
    \\
    &= 
    \sum^G_{g=1}
    \left[
        n_g(A_l) \log \hat{\theta}_g(A)
        +
        n_g(A_r) \log (1-\hat{\theta}_g(A))
    \right]
    \label{def of likelihood with MLE}
    ,
\end{align}
where the definition of $\hat{\theta}_g(A)$ (the MLE) and $r_i$ (the number of parameters) depend on which type of priors in Assumption \ref{assumption: sample size and true measures} is introduced by the state $i$:
\begin{align}
    \hat{\theta}_g(A) =
    \begin{cases}
    \frac{n_g(A_l)}{n_g(A)} & \text{(Prior A)}, \\
    \frac{n(A_l)}{n(A)} & \text{(Prior B)}, \\    
    \frac{P_g(A_l)}{P_g(A)} & \text{(Prior C)},
    \end{cases}
    \ \ \ 
    r_i =
    \begin{cases}
    G & \text{(Prior A)}, \\
    1 & \text{(Prior B)}, \\    
    0 & \text{(Prior C)}.
    \end{cases}   
    \label{def of MLE}
\end{align}
Notice that, for Prior C, the constant $c(A)$ and the true measures $P_g$ must satisfy \[
    c(A) = P_g(A_l \mid A) 
\]
for every $g = 1,\dots,G$ because if this does not hold, the state $i$ is not included in the set of feasible states $\tau(A \mid T)$.

Since $i \in \tau(A \mid T)$, the law of large numbers gives
$
    \hat{\theta}_g(A) \xrightarrow{p} P_g(A_l \mid A)
$.
Hence, we obtain the limit of $\hat{l}_A(i,T) / n$ as
\begin{align*}
    \frac{\hat{l}_A(i,T)}{n}
    &=
    \sum^G_{g=1}
    \frac{n_g(\Omega)}{n(\Omega)} 
    \frac{n_g(A)}{n_g(\Omega)}
    \left[
        \frac{n_g(A_l)}{n_g(A)}
        \log \hat{\theta}_g(A)
        +
        \frac{n_g(A_r)}{n_g(A)}
        \log (1-\hat{\theta}_g(A))      
    \right] \\
    &\xrightarrow{p}
    \sum^G_{g=1}
    \zeta_g P_g(A)
    \left[
        P_g(A_l \mid A) \log P_g(A_l \mid A)
        +
        P_g(A_r \mid A) \log P_g(A_r \mid A)        
    \right].
\end{align*}
\qed

\begin{lem}
\label{lem: comparison of log ML}
For $T \in \T^K$, $A \in \mathcal{N}(T)$, $i \in \tau(A \mid T)$ and $j \in \{1,\dots,I\}$, we have 
\begin{align*}
    \frac{\log M_i(A \mid T) - \log M_j(A \mid T)}{n} \xrightarrow{p} c_{i,j},
\end{align*}
where $c_{i,j} = 0$ if $j \in \tau(A \mid T)$ and $c_{i,j}>0$ if $j \in \{1,\dots,I\} \setminus \tau(A \mid T)$.
\end{lem}
\noindent 
(Proof)
If $j \in \tau(A \mid T)$, obtaining the result
\begin{align*}
    \frac{\log M_i(A \mid T) - \log M_j(A \mid T)}{n} \xrightarrow{p} 0
\end{align*}
is straightforward from the proof of Proposition \ref{lem: asymptotic behavior for a single node}. 
Hence we consider the case of $j \in \{1,\dots,I\} \setminus \tau(A \mid T)$.
Under the state $j$, for every $g$, the estimator $\hat{\theta}_g(A)$ is defined as in (\ref{def of MLE}), and there exists $C_g \in (0,1)$ such that $\hat{\theta}_g(A) \xrightarrow{p} C_g$.
By the definition of $\tau(A \mid T)$, there exists $g^*$ such that $C_{g^*} \neq P_{g^*}(A_l \mid A)$.
As in the proof of Proposition \ref{lem: asymptotic behavior for a single node}, for the difference of the marginal likelihoods, we obtain
\begin{align*}
    &\frac{\log M_i(A \mid T) - \log M_j(A \mid T)}{n}
    \xrightarrow{p}
    \sum^G_{g=1}
    \zeta_g P_g(A)
    \Lambda_{g}, \\
    &\Lambda_{g}
    =
    P_g(A_l \mid A) 
    \log \frac{P_g(A_l \mid A)}{C_g}
    +
    P_g(A_r \mid A) 
    \log \frac{P_g(A_r \mid A)}{1-C_g}.
\end{align*}
Because $\Lambda_g$ is the KL divergence of the two discrete distributions, $\Lambda_g \geq 0$ for all $g$ and $\Lambda_{g^*} > 0$.
By Assumption \ref{assumption: sample size and true measures}, this result implies that
\begin{align*}
\sum^G_{g=1}
    \zeta_g P_g(A)
    \Lambda_{g} > 0.
\end{align*}
\qed

\subsubsection*{Proof of Theorem~\ref{thm: characterizign the post of trees} \textcolor{black}{(non-informative prior)} and Theorem~\ref{thm: posterior of state variables} }
\textcolor{black}{
This section provides the proofs of Theorem~\ref{thm: characterizign the post of trees} under the assumption that the prior of the location variables is independent of the sample size, that is, $\eta = 0$ in Eq~\eqref{prior of L}. 
The general case is discussed in the next section. 
}

In this proof, we modify the notation for the marginal likelihood defined in Eq.~\eqref{def of MA} and use $M_i(A\mid T)$ to represent the likelihood on $A$ of the tree $T$ under the $i$th state to reflect its dependency on the tree structure.

 Let $T \in \T^K$ and $\sV$ denote a set of a combination of the states for all of the non-leaf nodes of $T$.
 Notice that an element of $\bV$ does not need to satisfy $P(\bV = \bv) > 0$, where $\bV$ is the totality of the state variables.
 In the following proof, for $\bv \in \sV$, $\bv(A)$ denotes a state on a node $A$.
Let $l (\bv, T)$ denote the log of the joint likelihood function
\begin{align}
     &l(\bv, T) = \log P(\x_n \mid T, \bv) 
     = \sum_{A \in \mathcal{N}(T)} l_A (\bv(A), T)
     + \sum_{A\in \mathcal{L}(T)} \log \mu(\x_n(A) \mid A),
     \label{non-asymptotic decomposition of the likelihood}
\end{align}
where $l_A (\bv(A), T) = \log M_A(\bv(A) \mid T)$.
By \cite{schwarz1978estimating}, this likelihood $l_A$ has the following expression
\begin{align*}
    l_A(\bv(A), T)
    &=
   \hat{l}_{A}(\bv(A) , T) 
   -
   \frac{r_{\bv(A)}}{2} \log n (A)
   + \Op,
\end{align*}
where $\hat{l}_{A}$ and $r_{i}$ are defined in (\ref{def of likelihood with MLE}).
Let $\bar{\bv} \in \sV$ be a collection of states such that, for all $A \in \mathcal{N}(T)$, $\theta_g(A)$ is fixed to $\mu(A_l)/\mu(A)$.
For $\bar{\bv}$, we have
\begin{align*}
    l_A(\bar{\bv}, T)
    &=
   \sum_{A \in \mathcal{N}(T)}
   \left\{
   n(A_l) \log \left(
        \frac{\mu(A_l)}{\mu(A)}
   \right)
   + 
   n(A_r) \log \left(
        \frac{\mu(A_r)}{\mu(A)}
   \right)
   \right\}
   +
   \sum_{A\in \mathcal{L}(T)} \log \mu(\x_n(A) \mid A)\\
   &= \log \mu(\x) = 0.
\end{align*}
Hence $l(\bv, T)$ is rewritten as
\begin{align*}
    &l(\bv, T) =  l(\bv, T) - l(\bar{\bv}, T) =
    \sum_{A \in \mathcal{N}(T)} 
    \left\{
        \hat{l}_{A}(\bv(A) , T) - \hat{l}_{A}(\bar{\bv}(A) , T)
    \right\}
   -
   \frac{C(\bv)}{2} \log n
   + \Op.
\end{align*}
For the part inside of the braces, when $\bv$ is replaced with $\bv_T \in \sV_T$, where $\sV_T = \{\bv: \bv(A) \in \tau(A \mid T) \text{ for all }
A \in \mathcal{N}(T) \}$, the definition of $\hat{l}_A$ gives
\begin{align*}
    \frac{\hat{l}(\bv_T(A), T) - \hat{l}(\bar{\bv}(A), T)}{n} 
    &=
    \sum^G_{g=1} \frac{n_g(A)}{n} 
    \left[
        \frac{n_g(A_l)}{n_g(A)} \log \frac{\hat{\theta}_g(A)}{\mu(A_l \mid A)}
        + \frac{n_g(A_r)}{n_g(A)} \log \frac{1- \hat{\theta}_g(A)}{\mu(A_r \mid A)}
    \right]\\
    &
    \xrightarrow{p}
    \sum^G_{g=1} \zeta_g P_g(A)
    \left[
        P_g(A_l \mid A) \log \frac{P_g(A_l \mid A)}{\mu(A_l \mid A)} 
        + P_g(A_r \mid A) \log \frac{P_g(A_r \mid A)}{\mu(A_r \mid A)} 
    \right].
\end{align*}
For all $A \in \mathcal{L}(T)$, there exists an unique sequence of nodes
\begin{align}
    \Omega = B_{A,0} \supset B_{A,1} \supset \cdots \supset B_{A,K} = A, \label{path of nodes}
\end{align}
where $B_{A,k} \in T$ ($k=0,\dots,K$) is a node in the $k$th level.
With this sequence, we obtain the limit of the scaled log-likelihood as
\begin{align}
    &\frac{l(\bv_T, T)}{n} \nonumber \\
    &\xrightarrow{p} \sum_{A \in \mathcal{N}(T)} \sum^G_{g=1} \zeta_g P_g(A)
    \left[
        P_g(A_l \mid A) \log \frac{P_g(A_l \mid A)}{\mu(A_l \mid A)} 
        + P_g(A_r \mid A) \log \frac{P_g(A_r \mid A)}{\mu(A_r \mid A)} 
    \right] \nonumber \\
    &= \sum^G_{g=1} \zeta_g\sum_{A \in \mathcal{N}(T)}  
    \left[
        P_g(A_l) \log \frac{P_g(A_l \mid A)}{\mu(A_l \mid A)} 
        + P_g(A_r) \log \frac{P_g(A_r \mid A)}{\mu(A_r \mid A)} 
    \right] \nonumber \\
    &=
    \sum^G_{g=1} \zeta_g \sum_{A \in \mathcal{L}(T)} P_g(A)
    \left[
        \log \frac{P_g(B_{A,1} \mid B_{A,0})}{\mu(B_{A,1} \mid B_{A,0})}
        + \cdots + 
        \log \frac{P_g(B_{A,K} \mid B_{A,K-1})}{\mu(B_{A,K} \mid B_{A,K-1})}
    \right] \nonumber \\
    &= \sum^G_{g=1} \zeta_g \sum_{A \in \mathcal{L}(T)} P_g(A) \log \frac{P_g(A)}{\mu(A)}
    =
    \sum^G_{g=1} \zeta_g KL(P_g |_T || \mu). 
    \label{limit of the best state variables}
\end{align}
Because $P_g|_T$ admits the density function
\begin{align*}
    p_g|_T(x)
    =
    \sum_{A \in \mathcal{L}(T)}
    \1_A (x) \frac{P_g(A)}{\mu(A)},\ x \in \Omega
\end{align*}
the KL divergence in (\ref{limit of the best state variables}) is rewritten as
\begin{align*}
    KL(P_g|_{T} || \mu) &= \sum_{A \in \mathcal{L}(T)} P_g(A) \log \frac{P_g(A)}{\mu(A)} \\
    &= \int p_g \sum_{A \in \mathcal{L}(T)} \mathbf{1}_A 
    \log \frac{P_g(A)}{\mu(A)} d \mu \\
    &= 
    \int p_g \log p_g |_{T} d\mu \\
    &= \int p_g(x) \log \frac{p_g(x)}{\mu(x)} d \mu(x)
    - \int p_g \log \frac{p_g}{p_g|_{T}} d \mu\\
    &= KL(P_g || \mu) - KL(P_g || P_g |_T).
\end{align*}
Because $KL(P_g || \mu)$ is independent of $T$, we obtain another expression of $\T^K_M$ in (\ref{def of TKM}) as 
\begin{align*}
\T^K_M &= 
\argmax_{T \in \T^K} 
\sum^G_{g=1}
\zeta_g
KL(P_g|_{T} || \mu)
\end{align*}
By Lemma \ref{lem: comparison of log ML} and (\ref{non-asymptotic decomposition of the likelihood}), for $\bv \in \sV \setminus \sV_T$,  we can show that
\begin{align}
    \plim_{n \to \infty}
    \frac{l(\bv, T) - l(\bv_T, T)}{n}
    =
    \plim_{n \to \infty}
    \sum_{A \in \mathcal{N}(T)}
    \frac{l_A(\bv, T) - l_A(\bv_T, T)}{n}
    > 0
    \label{plim of the difference of l},
\end{align}
and $\plim _{n \to \infty} l(\bv, T)/n$ exists.
Hence, for $T_M \in \T^K_M$, $\bv' \in \sV_{T_M}$, $T \in \T^K \setminus \T^K_M$ and $\bv \in \sV$, we have
\begin{align*}
    \plim_{n \to \infty}
    \frac{l(\bv', T_M) - l(\bv, T)}{n}
    \geq 
    \sum^G_{g=1} \zeta_g KL(P_g |_{T_M} || \mu)
    - 
    \sum^G_{g=1} \zeta_g KL(P_g |_T || \mu)
    > 0.
\end{align*}
Hence, for such $\T_M$ and $\bv'$, we obtain 
\begin{align*}
    \frac{P(\x_n \mid T)}{P(\x_n \mid T_M)}
    =
    \frac{ P(T)
    \sum_{\bv \in \sV} \exp(l(\bv, T)) P(\bv)}{ P(T_M) \sum_{\bv \in \sV} \exp(l(\bv, T_M)) P(\bv)}
    \leq
    \sum_{\bv \in \sV} 
    \frac{ \exp( l(\bv, T)) P(T) P(\bv)}{ \exp( l(\bv', T_M)) P(T_M) P(\bv')}    
    \xrightarrow{p} 0.
\end{align*}
This result implies $p(T \in \T^K_M \mid \x_n ) \xrightarrow{p} 1$, which completes the proof of Theorem \ref{thm: characterizign the post of trees}.

To prove Theorem \ref{thm: posterior of state variables}, we fix $T \in \T^K$ and define a set $\sS_T$ as
\begin{align*}
    \sS_T
    =
    \left\{
        \bv \in \sV_T 
        \mid 
        \bv \in 
        \argmin_{\bv' \in \sV_T }
        C(\bv' )
    \right\}.
\end{align*}
Then we want to show $P(\bV \in \sS_T \mid T, \x_n) \xrightarrow{p} 1$.
The result (\ref{plim of the difference of l}) implies 
\begin{align*}
    p(\bV \in \sV_T \mid T, \x_n)
    \xrightarrow{p}
    1,
\end{align*}
so we only need to compare the elements of $\sV_T$.
Let $\bv \in \sV_T \setminus \sS_T$ and $\bv' \in \sS_T$.
For the difference of the log likelihoods, we have
\begin{align*}
    l(\bv', T)  - l(\bv,T)
    &=
    \sum_{A \in \mathcal{N}(T)}
    \left[
        \hat{l}_A(\bv'(A), T) - \hat{l}_A(\bv(A), T)
    \right]
    +
    \frac{C(\bv) - C(\bv')}{2} \log n
    +
    \Op,
\end{align*}
where $\hat{l}_A$ and $C$ is defined in (\ref{def of MLE}) and (\ref{def: complexity}), respectively.
If $\bv(A)$ and $\bv'(A)$ introduce the same type of the prior (e.g., Prior A and Prior A), because the corresponding estimators $\hat{\theta}_g(A)$ have the same form,
\begin{align*}
    \hat{l}_A(\bv'(A), T) - \hat{l}_A(\bv(A), T) = 0.
\end{align*}
On the other hand, if $\bv(A)$ and $\bv'(A)$ introduce different types of the prior (e.g., Prior A and Prior B), because they are the maximized log-likelihood under the two nested hypotheses,
\begin{align*}
    -2 [\hat{l}_A(\bv'(A), T) - \hat{l}_A(\bv(A), T)]
\end{align*}
weakly converges to the $\chi^2$ distribution \citep{wilks1938large}.
Hence, we obtain
\begin{align*}
     \frac{l(\bv', T)  - l(\bv,T)}{\log n}
    \xrightarrow{p}  
    \frac{C(\bv) - C(\bv')}{2} > 0,
\end{align*}
which implies $P(\bV \in \sS_T\mid T, \x_n) \xrightarrow{p} 1$.
\qed

\subsubsection*{\textcolor{black}{Proof of Theorem~\ref{thm: characterizign the post of trees}  (informative prior) }}
\textcolor{black}{
   This section provides the proofs of Theorem~\ref{thm: characterizign the post of trees} for the case of the informative prior ($\eta > 0$).
    In the proof, we often use the results provided in the previous section on the non-informative prior and use the same notations. 
}

\textcolor{black}{
  We first discuss the asymptotic behavior of the prior of trees, which is in this case dependent on the sample size $n$.
  We only need to focus on the prior of the location variables $L$ since this is only the component that depends on the sample size as given in Eq~\eqref{prior of L}.
    By the fact that 
    \begin{align}
        \frac{n(A)}{n}
        \xrightarrow{p}
        \sum^G_{g=1}
        \zeta_g
        P_g(A) 
        \label{eq: limit of n(A)/n}
    \end{align}
    and Lemma \ref{lem: limit of log of sum}, we obtain the limit 
    \begin{align*}
        \frac{\log P(L(A) = \mu(A_l)/\mu(A))}{n}
        \xrightarrow{p}
        -
        \eta
        \sum^G_{g=1}
        \zeta_g
        P_g(A)
        f
        \left(
        \left|
            \frac{\mu(A_l)}{\mu(A)}
            -
            0.5
        \right|
        \right).
    \end{align*}
   Since given $T \in \T^K$, the location variables are all independent and the prior of the dimension variables is independent of the sample size $n$, the log-tree prior $l(T) = \log P(T)$ has a limit as follows:
\[
    \frac{l(T)}{n}
    \xrightarrow{p}
    - \eta 
    \sum^G_{g=1}
        \zeta_g
    B_g(T),
\]
which is the penalty term introduced in Theorem \ref{thm: characterizign the post of trees}.
}

\textcolor{black}{
We next define $\psi(T)$ as
\[
    \psi(T)
    =
    \sum^G_{g=1}
    \zeta_g 
    \left\{
        KL(P_g|_T || \mu) - \eta B_g(T)
    \right\}.
\]
By the the proof of Theorem \ref{thm: characterizign the post of trees} for the non-informative case, this is a limit under $\bv \in \sV_T$ (this $\sV_T$, and $\sV$, a collection of possible combination of the state variables on the tree, are defined in the proof for the non-informative case), that is,
\[
    \frac{l(T) + l(\bv, T)}{n}
    \xrightarrow{p} \psi(T)
\]
as $n \to \infty$.
On the other hand, for $\bv \in \sV \setminus \sV_T$, the limit exists and 
\[
    \plim \frac{l(T) + l(\bv, T)}{n} < \psi(T).
\]
As discussed in the proof of Theorem \ref{thm: characterizign the post of trees}, we have the equivalence
\[
    \argmax_{T \in \T^K}
    \sum^G_{g=1}
    \zeta_g 
    KL(P_g|_T || \mu)
    =
    \argmin_{T \in \T^K}
    \sum^G_{g=1}
    \zeta_g 
    KL(P_g || P_g |_T),
\]
which implies that 
\[
    T^K_M
    =
    \argmax_{T \in \T^K}
    \psi(T).
\]
Hence, for $T_M \in \T^K_M$, $\bv' \in \sV_{T_M}$, $T \in \T^K \setminus \T^K_M$, and $\bv \in \sV$, we have
\begin{align*}
    \plim_{n \to \infty}
    \frac{(l(T_M) + l(\bv', T_M) ) - (l(T) + l(\bv, T) )}
    {n}
    \geq
    \psi(T_M) - \psi(T) > 0.
\end{align*}
Hence, for such $T, \T_M$ and $\bv'$, it follows that
\begin{align*}
    \frac{P(T, \x_n)}{P(T_M, \x_n)}
    &=
    \frac{P(T) P(\x_n \mid T)}{P(T_M) P(\x_n \mid T_M)}
    =
    \frac{P(T) \sum_{\bv \in \sV}
        P(\bv) \exp(l(\bv, T))}{
        P(T_M) \sum_{\bv \in \sV}
        P(\bv) \exp(l(\bv, T_M))
    }\\
    &\leq 
    \sum_{\bv \in \sV}
    \frac{\exp(l(T) + l(\bv, T)) P(\bv)}{\exp(l(T_M) + l(\bv', T_M)) P(\bv')}
    \xrightarrow{p} 0.
\end{align*}
Therefore, $P(T \in \T^K_M \mid \x_n) \xrightarrow{p} 1$.
\qed
}

\section{Consistency for the MRS model}
\label{supp: Consistency for the MRS model}
To describe the consistency, for a possible node $A$, we define a variable $Z(A)$ as follows:
\begin{align*}
Z(A) =
    \begin{cases}
        1 &
        \text{ if } V(A) = 1,\\
        0 & 
        \text{ if } V(A)\in \{2,3\}.
    \end{cases}
\end{align*}
Hence, $\theta_1(A) = \theta_2(A)$ if $Z(A) = 0$ and $\theta_1(A) \neq \theta_2(A)$ with probability one if $Z(A) = 1$.
Then we can obtain the following consistency result.
\begin{cor}
     \label{cor: consistency for MRS}
     Let $\bZ= \{\bZ(A)\}_{A \in \mathcal{N}(T)}$ and $\bz = \{\bz(A)\}_{A \in \mathcal{N}(T)}$ be a collection of $Z(A)$ on $T \in \T^K$ and one of its realizations, respectively.
     If $P(\mathbf{Z} = \mathbf{z}) > 0$ for any possible $\mathbf{z}$, then
    \begin{align*}
         P
        \left(
             Z(A) =  
             \1_{\{
                 P_1(A_l \mid A) \neq P_2(A_l \mid A)
             \}}
             \text{ for all }
             A \in \mathcal{N}(T)
             \mid
             T, \x_n
         \right)
         \xrightarrow{p} 1
         ,
     \end{align*}
     where $\1$ is the indicator function, and 
     \begin{align*}
       P
        \left(
             Z(A) =  
             \1_{\{
                 P_1(A_l \mid A) \neq P_2(A_l \mid A)
             \}}
             \text{ for all }
             A \in \mathcal{N}(T)
             \mid
             \x_n
         \right)
         \xrightarrow{p} 1,
     \end{align*}
     where $T$ is random.
 \end{cor}
\noindent
(Proof)
In this case, $\sV_T$ in Theorem \ref{thm: posterior of state variables} is written as
\begin{align*}
    \sV_T =
    \left\{
        \bv \mid 
        \bv(A) = 1
        \text{ if }
        P_1(A_l \mid A) \neq P_2(A_l \mid A)
    \right\}.
\end{align*}
We additionally define $\tilde{\sV}_T$ as
\begin{align*}
\tilde{\sV}_T
=
    \left\{
        \bv \mid 
        \bv(A) = 2
        \text{ if }
        P_1(A_l \mid A) = P_2(A_l \mid A)
    \right\}.
\end{align*}
Then, under the condition that $\bv \in \sV_T$, the complexity $C(\bv)$ is minimized if and only if $\bv \in \tilde{\sV}_T$.
Hence, by Theorem \ref{thm: posterior of state variables} we obtain
\begin{align*}
    &         P
        \left(
             Z(A) =  
             \1_{\{
                 P_1(A_l \mid A) \neq P_2(A_l \mid A)
             \}}
             \text{ for all }
             A \in \mathcal{N}(T)
             \mid
             T, \x_n
         \right)\\
    &=
    P
    \left(
        \bV \in 
        \sV_T
        \cap 
        \tilde{\sV}_T
        \mid
        T, \x_n
    \right)\xrightarrow{p} 1.
\end{align*}
We can show the second result by using Theorem \ref{thm: characterizign the post of trees} as follows:
\begin{align*}
    P
    \left(
        \bV \in 
        \sV_T
        \cap 
        \tilde{\sV}_T
        \mid
        \x_n
    \right)
    &=
    \sum_{T \in \T^K}
    P
    \left(
        \bV \in 
        \sV_T
        \cap 
        \tilde{\sV}_T
        \mid
        T, \x_n
    \right)
    P(T \mid \x_n)\\
    &\geq
    \argmin_{T_M \in \T^K_M}
    P
    \left(
        \bV \in 
        \sV_T
        \cap 
        \tilde{\sV}_T
        \mid
        T_M, \x_n
    \right)
    P(T \in \T^K_M \mid \x_n)\\
    & \xrightarrow{p} 1.
\end{align*}
\qed

\section{Numerical evaluation of the density estimators under different complexity of data generating processes}
\label{sec: Additional experiment}

\textcolor{black}{
In this section, we observe the behavior of our proposed HMPT algorithm in density estimation in a case where observed points concentrate around certain points or the boundary of the sample space, which occurs in real data analysis in many cases. 
We consider the following scenario: the variables $X_j$ ($j=1,\dots,d$) independently follow the beta distribution 
\[
    {\rm Beta}(2^a, 2^a),
\]
where the parameter $a$ takes values from $\mathbb{R}$.
When $a=0$, the data is uniformly distributed in the sample space $(0,1]^d$, and when $a$ is positive, the distribution concentrates on the central point $(0.5,0.5,...,0.5)$. When $a$ is negative, the distribution concentrates on the vertices of the hyper-cube $(0,1]^d$.
For the cases of $a<0$, many values generated from the beta distribution tend to be indistinguishable from 0 and 1, and we found it often made the behavior of our HMPT algorithm, which is designed for continuous distributions without mass points, unstable.
As such, for the negative $a$'s, we use an approximative distribution in which the conditional distributions on $(0,10^{-5}]$ and $(0.9-10^{-5},1.0]$ are replaced with the uniform.
}

\textcolor{black}{
In the estimation, we used the APT model as described in Section \ref{sec: experiments (density estimation)}. The number of particles is 1,000, and the maximum resolution is set to 10 or 15. 
The data is generated under $a=(-3,-2,-1,0,1,2,3)$, $n = 10^2, 10^3, 10^4$, and $d = 1,2,3,4$, and for each combination, 50 data sets under different random seeds are generated.
The performance is assessed based on $L_1$ distance between the estimated density (posterior mean) and the true distribution.
Since the $L_1$ distance is difficult to compute analytically, we use the Monte Carlo approximation with 10000 values generated from the true distributions.
}

\begin{figure}[H]
\centering
\begin{tabular}{c}
    \includegraphics[height=14cm]{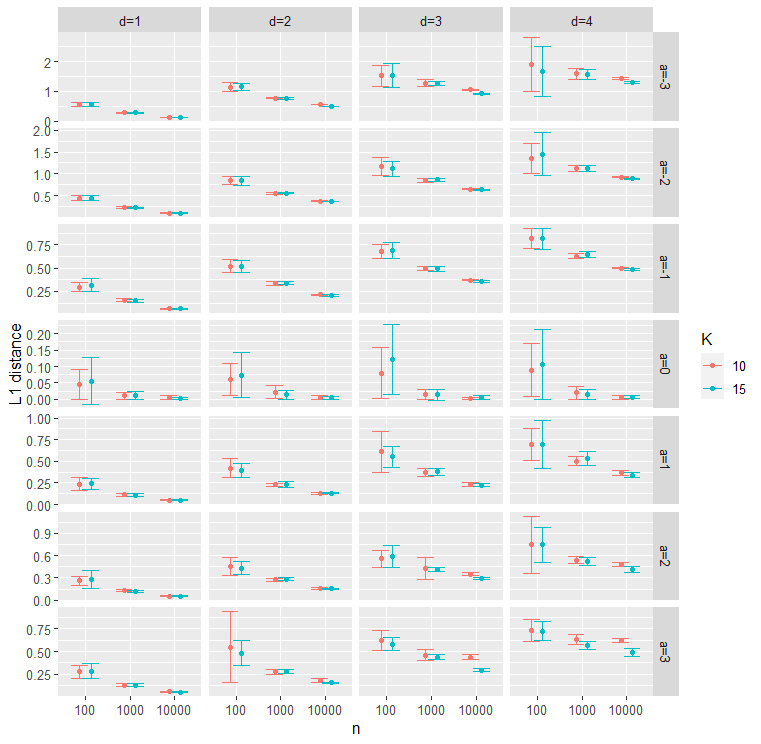}
\end{tabular}
\vspace{-0.5em}
\caption{ 
\textcolor{black}{
The average $L_1$ distance from the true distribution with error bars that indicate the standard deviation obtained under different sample size, dimensionality, and values of the parameter $a$.}
}
\label{fig: L1}
\end{figure}

\textcolor{black}{
Figure \ref{fig: L1} provides the average $L_1$ distance under the different settings. When the distribution is uniform ($a=0$), we can see that the $L_1$ distance is very small, and this can be understood as an advantage of using the APT model to learn the shrinkage level adaptively. On the other hand, especially when $a < 0$ and $d$ is large, the distance is larger for $a \neq 0$ though the performance is slightly improved if the maximum resolution $K$ is larger.
One possible reason to explain this result is that the APT model, with which we learn the smoothness of unknown densities, is not designed to capture such extremely spiky distributions. 
Thus the performance would be improved by modifying the prior distributions but this discussion is out of this research's scope because our main contribution is proposing the algorithm that works for many types of PT-based models.
}
 
\begin{figure}[H]
\centering
\begin{tabular}{c}
    \includegraphics[height=14cm]{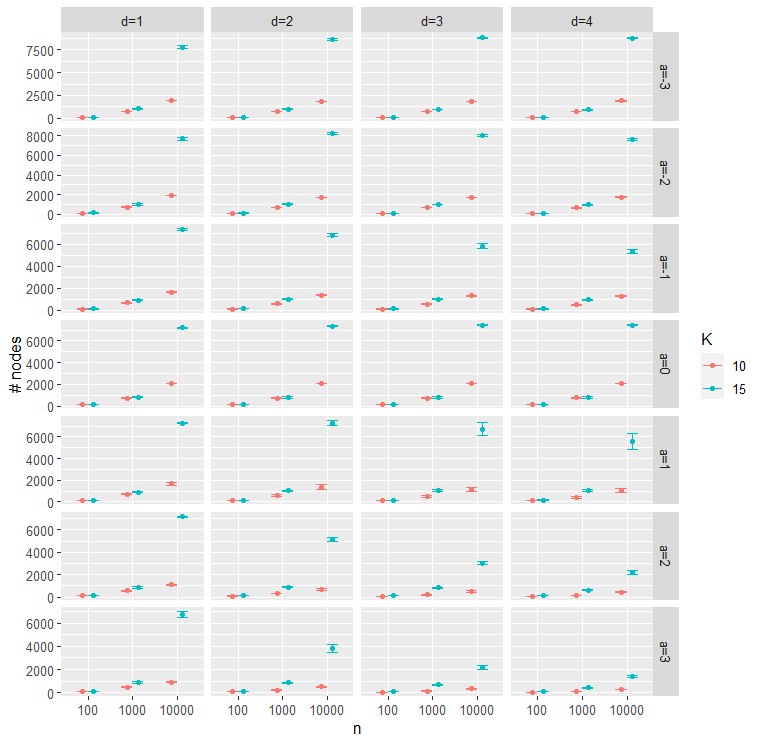}
\end{tabular}
\vspace{-0.5em}
\caption{ 
\textcolor{black}{
The average number of nodes in the MAP trees with error bars that indicate the standard deviation obtained under different sample size, dimensionality, and values of the parameter $a$.}
}
\label{fig: number of nodes}
\end{figure}

\textcolor{black}{
Figure \ref{fig: number of nodes} compares the number of nodes included in the MAP trees.
From this result, we can see that the number is larger for the larger maximum resolution $K=15$ as naturally expected but tends to be smaller when the parameter $a$ is positive. 
The latter phenomenon is explained by the rule that the SMC algorithm stops splitting nodes when the number of included observations is below the threshold such as 5.
When $a$ is positive, since the distribution concentrates around the central point, nodes with only a few observations tend to be generated in early levels and thus are no longer split, resulting in a smaller number of nodes on the tree. 
Hence this numerical result clarifies our proposed algorithm's tendency of ``ignoring'' nodes with only a few observations ``zooming up'' nodes including many observations.
}

\section{Details of the experiments}
\label{sec: Details of the experiemnts}

\subsection{Density estimation}
\subsubsection{Hyper-parameter settings}
For the HMPT model and the original APT model, we use the common settings following \cite{ma2017adaptive}.
The transition matrix for the latent states on each $A$ that characterize different smoothness levels of the density is given by
\vspace{-1em}
\begin{align*}
        \bxi_{i,i'}(A) 
        \propto \begin{cases} e^{\beta (i - i')} 
        & \text{ if } i \leq i', 
        \\
         0
        & \text{ if } i > i',
    \end{cases}
\end{align*}
\vspace{-2.5em}

\noindent
where $\beta = 0.1$ and the number of states $I = 5$.
For the $i(<I)$th state, given $V(A) = i$, the precision $\nu(A)$ follows the prior
\vspace{-1em}
\begin{align*}
    \log_{10} \nu(A) 
    &\sim
    \mathrm{Unif}(a(i), a(i+1)],
\end{align*}
\vspace{-2.5em}

\noindent
where $a(i) = L + (i-1)(U-L)/(I-1)$ with $L=-1$ and $U=4$.
For the $I$th state, $F_I = 1_\infty$, so $\nu(A)$ is fixed to $\infty$.
This is equivalent to stopping the partition and putting the conditional distribution $\mu(\cdot \mid A)$ on $A$.
In the computation, this uniform distribution is approximated by 5 evenly spaced grid points.

\subsubsection{Simulation scenarios in the two-dimensional cases}
The data sets are simulated from the following distributions.
\begin{enumerate}
    \item ``Blocks'': 
    \vspace{-3.7em}
    \begin{align*}
        &\frac{1}{3} \1_{[0.1, 0.45] \times [0.35, 0.9]}(x_1,x_2)
        + \frac{1}{3} \1_{[0.2, 0.8] \times [0.45, 0.5]}(x_1,x_2)\\
        &\hspace{5mm} + \frac{1}{3} \1_{[0.7, 0.9] \times [0.05, 0.6]}(x_1,x_2)
    \end{align*}
    \item ``Clusters'': 
    \begin{align*}
        &\frac{1}{10} {\rm Beta}(x_1 \mid 1,1) \times {\rm Beta}(x_2 \mid 1,1)
        + \frac{3}{10} {\rm Beta}(x_1 \mid 15,45)\times {\rm Beta}(x_2\mid 15,45) \\
        &\hspace{2.5mm}+ \frac{3}{10} {\rm Beta}(x_1 \mid 45,15)\times {\rm Beta}(x_2 \mid 22.5,37.5)
        + \frac{3}{10} {\rm Beta}(x_1 \mid 37.5,22.5)\times {\rm Beta}(x_2 \mid 45,15)
    \end{align*}
    \item ``Smooth'': 
        \vspace{-3.7em}
    \begin{align*}
        {\rm Beta}(x_1 \mid 10,20)  {\rm Beta}(x_2 \mid 10, 20)
    \end{align*}
\end{enumerate}

\subsubsection{Implementation details for the higher dimensional cases}
For the algorithm of the original APT (implemented by the {\tt apt} function in the {\tt PTT} package), in the first case with $d=6$, the maximum resolution is fixed to 9 because setting higher values lead to insufficient memory.
Also, because the {\tt apt} function does not scale if the dimension is beyond $d \approx 10$, in the second case with large $d$, we used the proposed SMC algorithm to carry out inference for the original APT model as well, which corresponds to setting $N_L = 2$. 

In the classical PT method, the Dirichlet prior is set to $\mathrm{Dir}(0.1k^2,...,0.1k^2)$ ($k$: the depth of the node), and the maximum depth is set to 15.
For the Dirichlet process Gaussian mixture model, we used the {\tt PYdensity} function in the R package {\tt BNPmix} \citep{corradin2021bnpmix}, in which the strength and discounting parameters are set to 10 and 0, respectively, the covariance matrices are all diagonal ({\tt model = "DLS"}), and the size of the burn-in period and the sampling is both set to 1000.
For the Gaussian kernel density estimation, it is implemented by the {\tt kde} function in the R package {\tt ks}. We set the bandwidth matrix by using {\tt Hpi.diag} function which selects the optimal diagonal matrix. 

\subsection{Two-group comparison}
\subsubsection{Hyper-parameter settings}
For the transition matrix $\bxi(A)$, we use the form proposed in \cite{soriano2017probabilistic} for incorporating multiple testing control
\vspace{-1em}
\begin{align*}
    \bxi(A)
    &=
    \left [
    \begin{array}{ccc}
        (1 - \rho)  \gamma&
        (1 - \rho) (1-  \gamma) &
        \rho \\
        (1 - \rho) \gamma 2^{-k}&
        (1 - \rho) (1- \gamma 2^{-k}) &
        \rho \\
        0 & 0 & 1
    \end{array}
    \right],
\end{align*}
\vspace{-1.5em}

\noindent
where $\gamma \in (0,1)$, and $\rho \in (0,1)$, and $k$ is the depth of $A$, and we set $(\gamma, \rho) = (0.3, 0.3)$ following recommendations in that paper.

\subsubsection{Simulation scenarios}
\begin{enumerate}
    \item ``Local location shift'': For $j=1,\dots,25$,
    \begin{align*}
        (X_{1,2(j-1)+1}, X_{1,2j})
        &\sim
        \frac{1}{3} N(\mu_1, \Sigma) + \sum^3_{l=2} \frac{1}{3} N(\mu_l, \Sigma), \\
        (X_{2,2(j-1)+1}, X_{2,2j})
        &\sim
        \frac{1}{3} N(\mu_1 + \delta_j, \Sigma) + \sum^3_{l=2} \frac{1}{3} N(\mu_l, \Sigma),
    \end{align*}
    where $\delta_j = -0.5$ for $j=1,\dots,5$ and 0 for $j=6,\dots,25$.
    \item ``Local dispersion difference'': For $j=1,\dots,25$,
    \begin{align*}
        (X_{1,2(j-1)+1}, X_{1,2j})
        &\sim
        \frac{1}{3} N(\mu_1, \Sigma) + \sum^3_{l=2} \frac{1}{3} N(\mu_l, \Sigma), \\
        (X_{2,2(j-1)+1}, X_{2,2j})
        &\sim
        \frac{1}{3} N(\mu_1, \Sigma + \Delta_j) + \sum^3_{l=2} \frac{1}{3} N(\mu_l, \Sigma),
    \end{align*}
    where $\Delta_j=-0.4$ for $j=1,\dots,5$ and 0 for $j=6,\dots,25$.
    \item ``Correlation'': For $j=1,\dots,25$,
    \begin{align*}
        (X_{1,2(j-1)+1}, X_{1,2j})
        &\sim
        N \left(
            \left[
                \begin{array}{c}
                    0  \\
                    0 
                \end{array}
            \right],
            \left[
                \begin{array}{cc}
                    1 & 0 \\
                    0 & 1
                \end{array}
            \right]
        \right) \\
        (X_{2,2(j-1)+1}, X_{2,2j})
        &\sim
        N \left(
            \left[
                \begin{array}{c}
                    0  \\
                    0 
                \end{array}
            \right],
            \left[
                \begin{array}{cc}
                    1 &  \delta_j \\
                    \delta_j & 1
                \end{array}
            \right]
        \right),
    \end{align*}
    where $\delta_j = 0.75$  for $j=1,\dots,5$ and $\delta_j = 0$  for $j=6,\dots,25$ .
\end{enumerate}
In the ``local location shift'' and ``local dispersion difference'', the parameters are 
\begin{align*}
    \mu_1 = (-2.5,1.0),\ \mu_2 = (1.0,-2.0),\ \mu_3 = (2.0, 2.5)
    ,\ \Sigma = 
    \left[
        \begin{array}{cc}
            0.5 &  0\\
            0 & 0.7
        \end{array}
    \right].
\end{align*}

\section{Additional figures}
\label{sec: additional figures}

\begin{figure}[htb]
\centering
\begin{tabular}{c}
    \includegraphics[height=5.5cm]{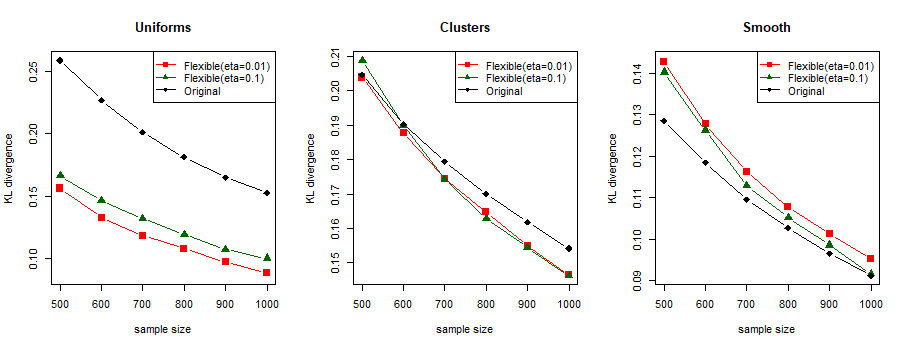}
\end{tabular}
\caption{The average KL divergences between the estimated density and the true density.}
\label{APT KL}
\end{figure}

\begin{figure}[htb]
\centering
\begin{tabular}{cc}
    \includegraphics[height=5.5cm]{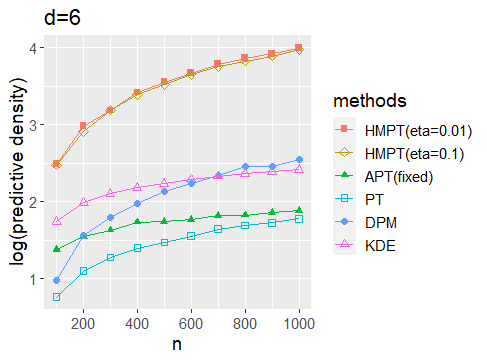}
    &
    \includegraphics[height=5.5cm]{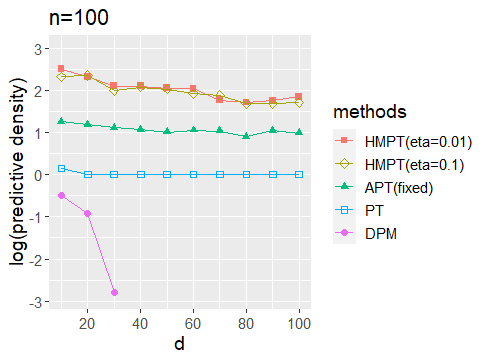}
\end{tabular}
\caption{
\textcolor{black}{The comparison of the predictive performance. Each point corresponds to the average of the predictive score in Eq.~\eqref{definition of D} based on 50 data sets. 
In the right plot, the predictive scores of the DPM model for the over 30 dimensional cases are below the displayed range. 
}
}
\label{fig: multi pred small}
\end{figure}

 \begin{figure}[tb]
 \centering
\begin{tabular}{c}
    \includegraphics[height=9cm]{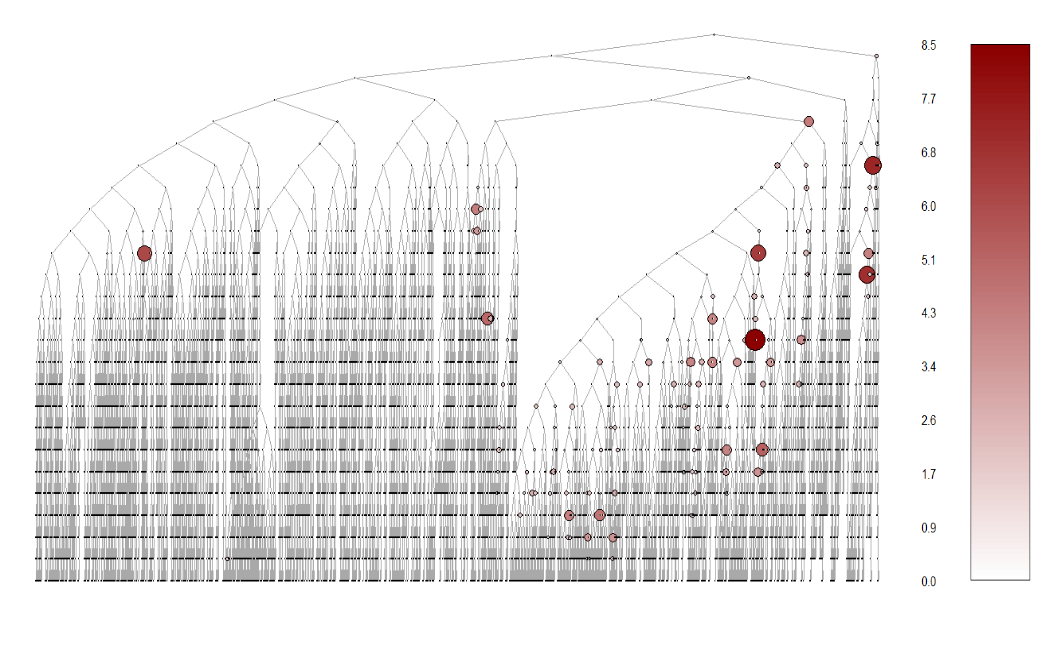}
\end{tabular}
  \caption{The MAP tree for the mass cytometry data set.  
  The size and the color indicate the estimated $\mathrm{eff}(A)$.
  Only the nodes with the sample size larger than 50 are drawn.
  Since there are a huge number of nodes, the information on the dimension is omitted in this figure.
  }
  \label{fig: tree for FC}
 \end{figure}

  \begin{figure}[htbp]
  \centering
  \includegraphics[width=1.0\hsize]{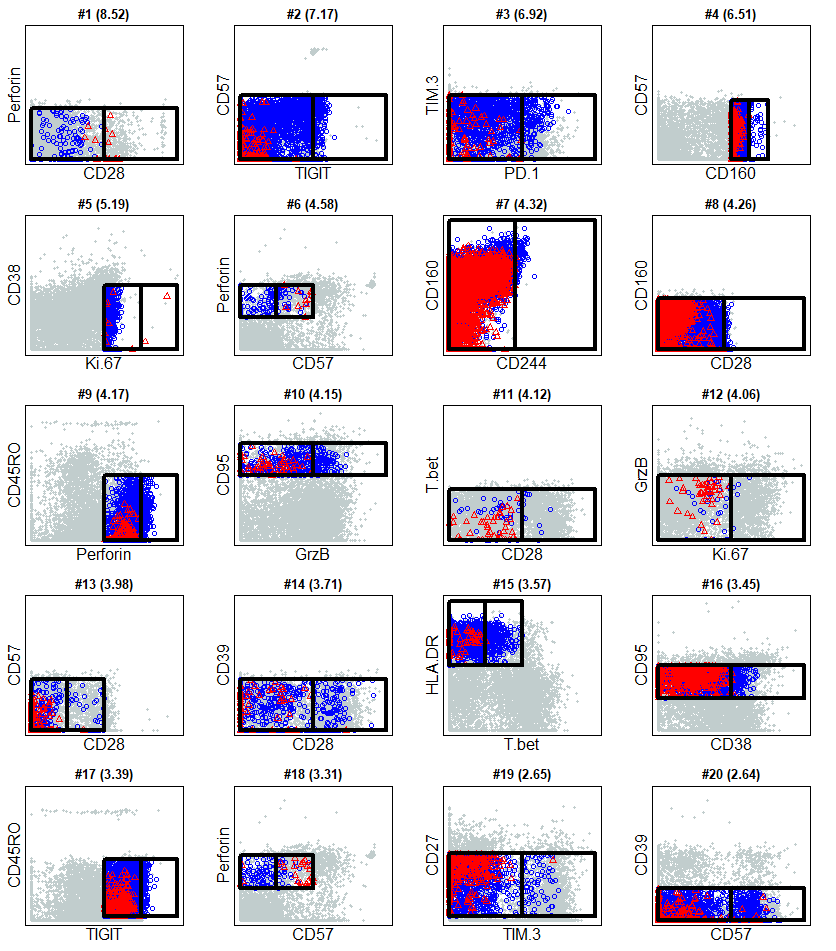}
  \caption{
The solid lines delineate the nodes with the highest values of $\mathrm{eff}(A)$ and their two children. The red triangle points and the blue circle points are the observations from the two samples in the node. The observations outside the node are in gray. In this figure, the nodes with $n_g(A)\geq 10$ $(g=1,2)$ are chosen.}
  \label{fig: node with large difference (FC)}
 \end{figure}

\end{document}